\input harvmac
\noblackbox

\ifx\answ\bigans
\magnification=1200\baselineskip=14pt plus 2pt minus 1pt
\else\baselineskip=16pt 
\fi
\def\ath{{\rm arctanh}}
\def\cf{{\it cf.\ }}
\def\ie{{\it i.e.\ }}
\def\eg{{\it e.g.\ }}
\def\eqq{{\it Eq.\ }}
\def\eqqs{{\it Eqs.\ }}
\def\th{\theta}

\def\eps{\epsilon}
\def\al{\alpha}
\def\de{\delta}
\def\be{\beta}
\def\Om{\Omega}
\def\om{\omega}
 
\def\ap{{a'}} 
\def\as{{a'a}}
\def\TH#1#2{\theta\left[#1\atop#2\right]}

\def\aarg{\lf(0,{it\o 2}\ri)}
\def\aargc{\lf(0,2il\ri)}
\def\UV{{\ $UV$\ }}

\def\at{\arctan(2U)}

\newif\ifnref
\def\rrr#1#2{\relax\ifnref\nref#1{#2}\else\ref#1{#2}\fi}
\def\ldf#1#2{\begingroup\obeylines
\gdef#1{\rrr{#1}{#2}}\endgroup\unskip}
\def\nrf#1{\nreftrue{#1}\nreffalse}
\def\multref#1#2#3{\nrf{#1#2#3}\refs{#1{--}#3}}

\def\doubref#1#2{\refs{{#1},{#2} }}
\def\threeref#1#2#3{\refs{{#1},{#2},{#3} }}
\def\fourref#1#2#3#4{\refs{{#1},{#2},{#3},{#4}}}

\nreffalse

\def\lref{\ldf}

\input epsf
\input psfig
\def\figin{\epsfcheck\figin}\def\figins{\epsfcheck\figins}
\def\epsfcheck{\ifx\epsfbox\UnDeFiNeD
\message{(NO epsf.tex, FIGURES WILL BE IGNORED)}
\gdef\figin##1{\vskip2in}\gdef\figins##1{\hskip.5in}
\else\message{(FIGURES WILL BE INCLUDED)}%
\gdef\figin##1{##1}\gdef\figins##1{##1}\fi}
\def\DefWarn#1{}
\def\figinsert{\goodbreak\midinsert}
\def\ifig#1#2#3{\DefWarn#1\xdef#1{fig.~\the\figno}
\writedef{#1\leftbracket fig.\noexpand~\the\figno}%
\figinsert\figin{\centerline{#3}}\medskip\centerline{\vbox{\baselineskip12pt
\advance\hsize by -1truein\noindent\footnotefont{\bf Fig.~\the\figno } #2}}
\bigskip\endinsert\global\advance\figno by1}

\def\appA{A}
\def\appB{B}\def\appBi{B.1.}\def\appBii{B.2.}
\def\appC{C}

\def\tilde{\widetilde}

\def\h {{1\over 2}}
\def\ov {\overline}
\def\o {\over}
\def\fc#1#2{{#1 \o #2}}

\def\IZ{ {\bf Z}}

\def\IR{ {\bf R}}


\def\br{\hfill\break}

\def\det {{\rm det}}

\def\lf {\left}
\def\ri {\right}
\def\ra {\rightarrow}

\def\re {{\rm Re}}
\def\im {{\rm Im}}
\def\p {\partial}

 \def\Fc {{\cal F}}
\def\Oc{{\cal O}}   
 \def\Vc {{\cal V}}\def\Rc {{\cal R}}


\lref\cvetic{M. Cvetic, G. Shiu and A.M. Uranga,
``Chiral four-dimensional N = 1 supersymmetric type IIA orientifolds from  intersecting 
D6-branes,''
Nucl.\ Phys.\ B {\bf 615}, 3 (2001)
[arXiv:hep-th/0107166].
}

\lref\ALL{M.~Berkooz, M.R.~Douglas and R.G.~Leigh, 
``Branes Intersecting at Angles,'' 
Nucl. Phys. B {\bf 480} (1996) 265, [arXiv:hep-th/9606139];\br
R.~Blumenhagen, L.~G\"orlich, B.~K\"ors and D.~L\"ust,
``Noncommutative compactifications of type I strings on tori with  
magnetic background flux,''
JHEP {\bf 0010}, 006 (2000)
[arXiv:hep-th/0007024];\br
C.~Angelantonj, I.~Antoniadis, E.~Dudas and A.~Sagnotti, 
``Type I Strings on Magnetized Orbifolds and Brane Transmutation'',
Phys. Lett. B {\bf 489} (2000) 223, [arXiv:hep-th/0007090];\br
C.~Angelantonj and A.~Sagnotti, 
``Type I Vacua and Brane Transmutation'', 
hep-th/0010279;\br
G.~Aldazabal, S.~Franco, L.E.~Ibanez, R.~Rabadan and A.M.~Uranga,
``D = 4 chiral string compactifications from intersecting branes,''
J.\ Math.\ Phys.\  {\bf 42}, 3103 (2001)
[arXiv:hep-th/0011073];\br
G.~Aldazabal, S.~Franco, L.E.~Ibanez, R.~Rabadan, and A.M.~Uranga,
``Intersecting Brane Worlds'', 
JHEP {\bf 0102} (2001) 047, [arXiv:hep-ph/0011132];\br
R.~Blumenhagen, B.~K\"ors and D.~L\"ust,
``Type I strings with F- and B-flux,''
JHEP {\bf 0102}, 030 (2001)
[arXiv:hep-th/0012156];\br
L.E.~Ibanez, F.~Marchesano, R.~Rabadan, 
``Getting just the Standard Model at Intersecting Branes'',
JHEP {\bf 0111} (2001) 002, [arXiv:hep-th/0105155];\br
S. F\"orste, G. Honecker and R. Schreyer,
``Orientifolds with branes at angles,''
JHEP {\bf 0106}, 004 (2001)
[arXiv:hep-th/0105208];\br
R.~Rabadan, 
``Branes at Angles, Torons, Stability and Supersymmetry'', 
Nucl.\ Phys.\ B {\bf 620} (2002) 152, [arXiv:hep-th/0107036];\br
R.~Blumenhagen, B.~K\"ors, D.~L\"ust and T.~Ott,
``The standard model from stable intersecting brane world orbifolds,''
Nucl.\ Phys.\ B {\bf 616}, 3 (2001)
[arXiv:hep-th/0107138];\br
M. Cvetic, G. Shiu and A.M. Uranga,
``Chiral four-dimensional N = 1 supersymmetric type IIA orientifolds from  intersecting 
D6-branes,''
Nucl.\ Phys.\ B {\bf 615}, 3 (2001)
[arXiv:hep-th/0107166];\br
D.~Bailin, G.V.~Kraniotis and A.~Love, 
``New Standard-like Models from Intersecting D4-Branes'',  
hep-th/0208103;
``Standard-like models from intersecting D5-branes'',  
hep-th/0210227;\br
D.~Cremades, L.E.~Ibanez and F.~Marchesano,
``SUSY quivers, intersecting branes and the modest hierarchy problem,''
JHEP {\bf 0207}, 009 (2002)
[arXiv:hep-th/0201205];\br
R.~Blumenhagen, B.~K\"ors and D.~L\"ust,
``Moduli stabilization for intersecting brane worlds in type 0' string  theory,''
Phys.\ Lett.\ B {\bf 532}, 141 (2002)
[arXiv:hep-th/0202024];\br
D.~Cremades, L.E.~Ibanez and F.~Marchesano, 
    ``Intersecting Brane Models of Particle Physics and the Higgs Mechanism'',
   JHEP {\bf 0207} (2002) 022,   [arXiv:hep-th/0203160];\br
C.~Kokorelis, 
``GUT Model Hierarchies from Intersecting Branes'',
JHEP {\bf 0208} (2002) 018, [arXiv:hep-th/0203187];
``Deformed Intersecting D6-Brane GUTS I'',
hep-th/0209202;
``Deformed Intersecting D6-Brane GUTS II'',
hep-th/0210200;\br
R.~Blumenhagen, V.~Braun, B.~K\"ors and D.~L\"ust,
``Orientifolds of K3 and Calabi-Yau manifolds with intersecting D-branes,''
JHEP {\bf 0207}, 026 (2002)
[arXiv:hep-th/0206038];
``The standard model on the quintic,''
hep-th/0210083;\br
A.M.~Uranga,
``Local models for intersecting brane worlds'', 
JHEP {\bf 0212} (2002) 058, [arXiv:hep-th/0208014];\br
G.~Pradisi, 
``Magnetized (Shift-)Orientifolds'', 
hep-th/0210088;\br
R.~Blumenhagen, L. G\"orlich and T. Ott, 
``Supersymmetric Intersecting Branes on the Type IIA $T^6/Z_4$ orientifold'',
JHEP {\bf 0301} (2003) 021, hep-th/0211059;\br
M. Cvetic, I. Papadimitriou and G. Shiu,
``Supersymmetric Three Family SU(5) Grand Unified Models from Type IIA Orientifolds 
with Intersecting D6-Branes'', 
hep-th/0212177.}

\lref\cveticaa{M. Cvetic, I. Papadimitriou and G. Shiu,
``Supersymmetric Three Family SU(5) Grand Unified Models from Type IIA Orientifolds 
with Intersecting D6-Branes'', 
hep-th/0212177}

\lref\bonnii{
S. F\"orste, G. Honecker and R. Schreyer,
``Orientifolds with branes at angles,''
JHEP {\bf 0106}, 004 (2001)
[arXiv:hep-th/0105208];\br
G. Honecker, PhD thesis, Bonn University 2002.}

\lref\bonni{
S. F\"orste, G. Honecker and R. Schreyer,
``Supersymmetric $\IZ_N \times \IZ_M$ orientifolds in 4D with D-branes at angles,''
Nucl.\ Phys.\ B {\bf 593}, 127 (2001)
[arXiv:hep-th/0008250];\br
R. Schreyer, PhD thesis, Bonn University 2001.}

\lref\berlinii{
R.~Blumenhagen, L.~G\"orlich and B.~K\"ors,
``Supersymmetric orientifolds in 6D with D-branes at angles,''
Nucl.\ Phys.\ B {\bf 569}, 209 (2000)
[arXiv:hep-th/9908130].
}

\lref\berlini{
R.~Blumenhagen, L.~G\"orlich and B.~K\"ors,
``Supersymmetric 4D orientifolds of type IIA with D6-branes at angles,''
JHEP {\bf 0001}, 040 (2000)
[arXiv:hep-th/9912204].
}

\lref\phencvetic{M.~Cvetic, P.~Langacker and G.~Shiu,
``Phenomenology of a three-family standard-like string model,''
Phys.\ Rev.\ D {\bf 66}, 066004 (2002)
[arXiv:hep-ph/0205252];\br
G.~Shiu and S.H.~Tye,
``TeV scale superstring and extra dimensions,''
Phys.\ Rev.\ D {\bf 58}, 106007 (1998)
[arXiv:hep-th/9805157].
}

\lref\wise{
J.~Gomis, T.~Mehen and M.B.~Wise,
``Quantum field theories with compact noncommutative extra dimensions,''
JHEP {\bf 0008}, 029 (2000)
[arXiv:hep-th/0006160];\br
Z.~Guralnik, R.C.~Helling, K.~Landsteiner and E.~Lopez,
``Perturbative instabilities on the non-commutative torus, Morita duality  and 
twisted boundary conditions,''
JHEP {\bf 0205}, 025 (2002)
[arXiv:hep-th/0204037].
}

\lref\LSii{D. L\"ust and S. Stieberger, to appear.}
\lref\progress{Work in progress.}

\lref\ABD{
I.~Antoniadis, C.~Bachas and E.~Dudas,
``Gauge couplings in four-dimensional type I string orbifolds,''
Nucl.\ Phys.\ B {\bf 560}, 93 (1999)
[arXiv:hep-th/9906039].
}

\lref\DKL{L.J.~Dixon, V.~Kaplunovsky and J.~Louis,
``Moduli Dependence Of String Loop Corrections To Gauge Coupling Constants,''
Nucl.\ Phys.\ B {\bf 355}, 649 (1991).}
\lref\other{I.~Antoniadis, K.S.~Narain and T.R.~Taylor,
``Higher genus string corrections to gauge couplings,''
Phys.\ Lett.\ B {\bf 267}, 37 (1991);\br
I.~Antoniadis, E.~Gava and K.S.~Narain,
``Moduli corrections to gauge and gravitational couplings in four-dimensional superstrings,''
Nucl.\ Phys.\ B {\bf 383}, 93 (1992)
[arXiv:hep-th/9204030];\br
I.~Antoniadis, E.~Gava, K.S.~Narain and T.R.~Taylor,
``Superstring threshold corrections to Yukawa couplings,''
Nucl.\ Phys.\ B {\bf 407}, 706 (1993)
[arXiv:hep-th/9212045];\br
P.~Mayr and S.~Stieberger,
``Threshold corrections to gauge couplings in orbifold compactifications,''
Nucl.\ Phys.\ B {\bf 407}, 725 (1993)
[arXiv:hep-th/9303017];
``Dilaton, antisymmetric tensor and gauge fields in string effective theories at the 
one loop level,''
Nucl.\ Phys.\ B {\bf 412}, 502 (1994)
[arXiv:hep-th/9304055];
``Moduli dependence of one loop gauge couplings in (0,2) compactifications,''
Phys.\ Lett.\ B {\bf 355}, 107 (1995)
[arXiv:hep-th/9504129];\br
E.~Kiritsis and C.~Kounnas,
``Infrared regularization of superstring theory and the one loop 
calculation of coupling constants,''
Nucl.\ Phys.\ B {\bf 442}, 472 (1995)
[arXiv:hep-th/9501020];\br
E.~Kiritsis, C.~Kounnas, P.M.~Petropoulos and J.~Rizos,
``Universality properties of N = 2 and N = 1 heterotic threshold  corrections,''
Nucl.\ Phys.\ B {\bf 483}, 141 (1997)
[arXiv:hep-th/9608034];\br
J.A.~Harvey and G.W.~Moore,
``Algebras, BPS States, and Strings,''
Nucl.\ Phys.\ B {\bf 463}, 315 (1996)
[arXiv:hep-th/9510182];\br
H.P. Nilles and S.~Stieberger,
``String unification, universal one-loop corrections and strongly coupled  heterotic string 
theory,''
Nucl.\ Phys.\ B {\bf 499}, 3 (1997)
[arXiv:hep-th/9702110];\br
S.~Stieberger,
``(0,2) heterotic gauge couplings and their M-theory origin,''
Nucl.\ Phys.\ B {\bf 541}, 109 (1999)
[arXiv:hep-th/9807124].
}

\lref\FS{K. Foerger and S. Stieberger,
``Higher derivative couplings and heterotic-type I duality in eight  dimensions,''
Nucl.\ Phys.\ B {\bf 559}, 277 (1999)
[arXiv:hep-th/9901020].
}

\lref\CIM{D.~Cremades, L.E.~Ibanez and F.~Marchesano,
``SUSY quivers, intersecting branes and the modest hierarchy problem,''
JHEP {\bf 0207}, 009 (2002)
[arXiv:hep-th/0201205].
}
\lref\BKLO{
R.~Blumenhagen, B.~K\"ors, D.~L\"ust and T.~Ott,
``The standard model from stable intersecting brane world orbifolds,''
Nucl.\ Phys.\ B {\bf 616}, 3 (2001)
[arXiv:hep-th/0107138].
}

\lref\BP{C.~Bachas and M.~Porrati,
``Pair Creation Of Open Strings In An Electric Field,''
Phys.\ Lett.\ B {\bf 296}, 77 (1992)
[arXiv:hep-th/9209032].}
\lref\BF{C.~Bachas and C.~Fabre,
``Threshold Effects in Open-String Theory,''
Nucl.\ Phys.\ B {\bf 476}, 418 (1996)
[arXiv:hep-th/9605028].
}

\lref\AFIRU{G.~Aldazabal, S.~Franco, L.E.~Ibanez, R.~Rabadan and A.M.~Uranga,
``D = 4 chiral string compactifications from intersecting branes,''
J.\ Math.\ Phys.\  {\bf 42}, 3103 (2001)
[arXiv:hep-th/0011073].
}

\lref\BGKL{
R.~Blumenhagen, L.~G\"orlich, B.~K\"ors and D.~L\"ust,
``Noncommutative compactifications of type I strings on tori with  
magnetic background flux,''
JHEP {\bf 0010}, 006 (2000)
[arXiv:hep-th/0007024].
}

\lref\IRAN{F.~Ardalan, H.~Arfaei and M.M.~Sheikh-Jabbari,
``Noncommutative geometry from strings and branes,''
JHEP {\bf 9902}, 016 (1999)
[arXiv:hep-th/9810072].
}

\lref\LNSW{W.~Lerche, B.E.~Nilsson, A.N.~Schellekens and N.P.~Warner,
``Anomaly Cancelling Terms From The Elliptic Genus,''
Nucl.\ Phys.\ B {\bf 299}, 91 (1988).
}

\lref\LSW{A.N. Schellekens and N.P.~Warner,
``Anomalies, Characters And Strings,''
Nucl.\ Phys.\ B {\bf 287}, 317 (1987);
``Anomalies And Modular Invariance In String Theory,''
Phys.\ Lett.\ B {\bf 177}, 317 (1986);\br
W.~Lerche,
``Elliptic Index And Superstring Effective Actions,''
Nucl.\ Phys.\ B {\bf 308}, 102 (1988);\br
W.~Lerche, A.N.~Schellekens and N.P.~Warner,
``Lattices And Strings,''
Phys.\ Rept.\  {\bf 177}, 1 (1989).
}

\lref\quintic{
R.~Blumenhagen, V.~Braun, B.~K\"ors and D.~L\"ust,
``Orientifolds of K3 and Calabi-Yau manifolds with intersecting D-branes,''
JHEP {\bf 0207}, 026 (2002)
[arXiv:hep-th/0206038];
``The standard model on the quintic,''
hep-th/0210083.
}

\lref\BlumenhagenEA{
R.~Blumenhagen, B.~K\"ors and D.~L\"ust,
``Type I strings with F- and B-flux,''
JHEP {\bf 0102}, 030 (2001)
[arXiv:hep-th/0012156].
}

\lref\BGKLnc{
R.~Blumenhagen, L.~G\"orlich, B.~K\"ors and D.~L\"ust,
``Asymmetric orbifolds, noncommutative geometry and type I string vacua,''
Nucl.\ Phys.\ B {\bf 582}, 44 (2000)
[arXiv:hep-th/0003024].
}

\lref\callan{A.~Abouelsaood, C.G.~Callan, C.R.~Nappi and S.A.~Yost,
``Open Strings In Background Gauge Fields,''
Nucl.\ Phys.\ B {\bf 280}, 599 (1987);
``String Loop Corrections To Beta Functions,''
Nucl.\ Phys.\ B {\bf 288}, 525 (1987).
}

\lref\tseytlin{E.S.~Fradkin and A.A.~Tseytlin,
``Nonlinear Electrodynamics From Quantized Strings,''
Phys.\ Lett.\ B {\bf 163}, 123 (1985);\br
R.R.~Metsaev and A.A.~Tseytlin,
``On Loop Corrections To String Theory Effective Actions,''
Nucl.\ Phys.\ B {\bf 298}, 109 (1988);\br
A.A.~Tseytlin,
``Vector Field Effective Action In The Open Superstring Theory,''
Nucl.\ Phys.\ B {\bf 276}, 391 (1986)
[Erratum-ibid.\ B {\bf 291}, 876 (1987)].
}

\lref\HM{J.A.~Harvey and G.W.~Moore,
``Algebras, BPS States, and Strings,''
Nucl.\ Phys.\ B {\bf 463}, 315 (1996)
[arXiv:hep-th/9510182].
}

\lref\LS{W.~Lerche and S.~Stieberger,
``Prepotential, mirror map and F-theory on K3,''
Adv.\ Theor.\ Math.\ Phys.\  {\bf 2}, 1105 (1998)
[Erratum-ibid.\  {\bf 3}, 1199 (1999)]
[arXiv:hep-th/9804176].
}

\lref\Grad{
I.S. Gradshteyn and I.M. Ryzhik,
``Table of Integrals, Series and Products'', Academic Press 1994.}

\lref\APO{T.M. Apostol
``Modular functions and Dirichlet series in number theory'', Springer 1997.}

\lref\FW{
T.~Friedmann and E.~Witten,
``Unification scale, proton decay, and manifolds of G(2) holonomy,''
hep-th/0211269.
}

\lref\STii{S.~Stieberger and T.R.~Taylor,
``Non-Abelian Born-Infeld action and type I - heterotic duality.  II: 
Nonrenormalization theorems,''
Nucl.\ Phys.\ B {\bf 648}, 3 (2003)
[arXiv:hep-th/0209064].
}

\lref\heterotic{
L.E.~Ibanez, D. L\"ust and G.G.~Ross,
``Gauge coupling running in minimal $SU(3) \times SU(2) \times U(1)$ superstring 
unification,''
Phys.\ Lett.\ B {\bf 272}, 251 (1991)
[arXiv:hep-th/9109053];\br
L.E.~Ibanez and D.~L\"ust,
``Duality anomaly cancellation, minimal string unification and the effective 
low-energy Lagrangian  of 4-D strings,''
Nucl.\ Phys.\ B {\bf 382}, 305 (1992)
[arXiv:hep-th/9202046];\br
P.~Mayr, H.P.~Nilles and S.~Stieberger,
``String unification and threshold corrections,''
Phys.\ Lett.\ B {\bf 317}, 53 (1993)
[arXiv:hep-th/9307171];\br
H.P. Nilles and S.~Stieberger,
``How to Reach the Correct $\sin^2\theta_W$ and $\alpha_S$ in String Theory,''
Phys.\ Lett.\ B {\bf 367}, 126 (1996)
[arXiv:hep-th/9510009];
``String unification, universal one-loop corrections and strongly coupled  heterotic string 
theory,''
Nucl.\ Phys.\ B {\bf 499}, 3 (1997)
[arXiv:hep-th/9702110].
}

\lref\BGO{R.~Blumenhagen, L. G\"orlich and T. Ott,
``Supersymmetric intersecting branes on the type IIA $T^6/\IZ_4$  orientifold,''
hep-th/0211059.}

\lref\GNS{D.M.~Ghilencea, H.P.~Nilles and S.~Stieberger,
``Divergences in Kaluza-Klein models and their string regularization,''
New J.\ Phys.\  {\bf 4}, 15 (2002)
[arXiv:hep-th/0108183].
}

\lref\LSquarter{
W.~Lerche and S.~Stieberger,
``1/4 BPS states and non-perturbative couplings in N = 4 string theories,''
Adv.\ Theor.\ Math.\ Phys.\  {\bf 3}, 1539 (1999)
[arXiv:hep-th/9907133].
}

\lref\morales{A.B.~Hammou and J.F.~Morales,
``Fivebrane instantons and higher derivative couplings in type I theory,''
Nucl.\ Phys.\ B {\bf 573}, 335 (2000)
[arXiv:hep-th/9910144].
}

\lref\KaplunovskyRP{
V.~S.~Kaplunovsky,
``One Loop Threshold Effects In String Unification,''
Nucl.\ Phys.\ B {\bf 307}, 145 (1988)
[Erratum-ibid.\ B {\bf 382}, 436 (1992)]
[arXiv:hep-th/9205068].
}

\Title{\vbox{\rightline{HU--EP--03/08} 
\rightline{\tt hep-th/0302221}}}
{\vbox{\centerline{Gauge Threshold Corrections in}
\bigskip\centerline{Intersecting Brane World Models}}}
\smallskip
\centerline{D. L\"ust and\ S. Stieberger}
\bigskip
\centerline{\it Institut f\"ur  Physik, Humboldt Universit\"at zu Berlin,}
\centerline{\it Invalidenstra\ss e 110, 10115 Berlin, FRG}

\bigskip\bigskip
\centerline{\bf Abstract}
\vskip .2in
\noindent
We calculate the one--loop corrections to gauge couplings in 
N=1 supersymmetric brane world models, which are realized in an type $IIA$ 
orbifold/orientifold background with several stacks of D6 branes wrapped on 3--cycles
with non--vanishing intersections.
Contributions arise from both N=1 and N=2 open string subsectors.
In contrast to what is known from ordinary orbifold theories, 
N=1 subsectors do give rise to moduli--dependent one--loop corrections.

\Date{}
\noindent

\goodbreak

\newsec{Introduction}

Intersecting brane world models \ALL\ 
have proven to be provide an interesting
framework of getting string compactifications with promising
phenomenological features. For the type IIA superstring these compactifications 
contain in particular several types of D6-branes which are wrapped
around 3-cycles of the internal space. Chiral fermions appear
at intersections of branes at angles.
The chiral
fermion spectrum is determined by the topological intersection numbers
of the relevant 3-cycles. 
Part of space-time supersymmetry is preserved if the D6-branes are wrapped
around supersymmetric (special lagrangian) 3-cycles, which are calibrated with
respect to the
same holomorphic 3-form as the O6-planes are.
This scenario was intensively investigated
in the context of toroidal type IIA orientifolds and also for Calabi-Yau
orientifolds, and in fact models with
spectrum identical  to the non-supersymmetric or supersymmetric Standard Model
could be explicitly constructed. In a T-dual respectively mirror symmetric
description
one is dealing with D9-branes with additional gauge fluxes turned on.
This can been seen very explicitly for the toroidal models after
performing the T-duality transformation with respect to three internal
directions. Then the various angles of the
D6-branes translate themselves into constant magnetic gauge fluxes
on the D9-branes, such that the internal torus becomes non--commutative.

In the present paper we turn to the question of computing one--loop gauge
threshold corrections in intersecting brane world models, which is
also very important from the phenomenological point of view.
Unlike what happens \eg in perturbative heterotic string vacua, the tree--level gauge 
couplings for the various gauge groups, arising from different stacks of branes, are not 
the same at the string scale.
They follow from dimensional reducing the Born--Infeld action of a $D6$--brane on a  
$3$--cycle of an internal six--torus $T_6$ and 
are essentially determined by the volume of the $3$--cycle. 
{\it E.g.} for a six--torus $T_6=\prod\limits_{j=1}^3 T_2^j$ and a 
special $3$--cycle embedded with the wrapping numbers $(n^j_a,m^j_a)$ w.r.t. to the 
two--tori $T_2^j$  the gauge couplings are given by\foot{The imaginary part of the standard $D=4$ 
dilaton field $S$ follows for $(n^j_a,m^j_a)=(1,0)$, \ie $\im(S)=g^{-2}_{\rm string}=
\fc{M^3_{\rm string}}{2\pi \lambda_{II}}\ R_1^1R_1^2R_1^3$.} \doubref\CIM\phencvetic
\eqn\treegauge{
g_{a,\ \rm tree}^{-2}=\fc{M^3_{\rm string}}{2\pi \lambda_{II}} \ \prod_{j=1}^3
\sqrt{(n^j_a)^2 (R^j_1)^2+(m^j_a)^2 (R^j_2)^2 + 2 n^j_a m_a^j  R_1^jR_2^j\cos\alpha^j}\ ,}
with the type $II$ coupling constant $\lambda_{II}$ (\cf the next section for more details).
Hence a priori
there is no unification of gauge couplings at the string scale (at string tree--level).
One--loop gauge threshold corrections $\Delta_a$ (to the gauge group $G_a$), which take 
into account Kaluza--Klein and winding states from the internal dimensions and 
the heavy string modes, may change this picture  \KaplunovskyRP. 
For certain regions in moduli 
space these corrections may become huge and thus have a substantial impact\foot{This effect 
has been thoroughly investigated for heterotic N=1 string vacua in 
\multref\DKL\other\heterotic\ and will
be discussed for the models under discussion elsewhere \progress.} on the 
unification scale.
One--loop gauge corrections  are very important quantities to probe the 
low--energy physics below the string scale as they change the running of the gauge couplings 
for scales $\mu$ below the string scale
according to the Georgi, Quinn and Weinberg evolution equations of ordinary field theories:
\eqn\weinberg{
\fc{1}{g^2_a(\mu)}=\fc{1}{g_{a,\ \rm tree}^2}+\fc{b_a}{16\pi^2}\ 
\ln\fc{M_{\rm string}^2}{\mu^2}+\Delta_a\ .}

In the type IIA picture with intersecting D6-branes these threshold
correction $\Delta_a$ will depend on the holomolgy classes on the 3-cycles 
(open string parameters) and also
on the closed string geometrical moduli.
In toroidal models these corrections will be given in terms of the
wrapping numbers $n_a^j,m_a^j$ and the radii $R_i^j$ of the torus,   
in analogy to their tree--level counterparts \treegauge, however in a more 
complicated way.
In the equivalent T-dual picture the threshold corrections will be a function of the 
open string magnetic gauge fluxes and of the geometrical moduli of the
dual compact space.
Since the gauge fluxes are directly related to the non-commutativity parameters
of the internal torus, we obtain in this way some interesting, new informations
for one--loop threshold corrections on non-commutative tori
in string theory, a discussion which extends recent results
on one--loop corrections on compact non-commutative spaces in the literature
\wise\ and will be further discussed in \LSii.

In this article we shall calculate the quantities $\Delta_a$ for 
a class of realistic brane world models realized through intersecting branes which are
wrapped on internal tori. We mainly focus on supersymmetric intersecting brane
models and leave the discussion of the non-supersymmetric case for a future
publication \LSii.
The main motivation to discuss supersymmetric models is, that in these
theories vacuum $R$--tadpoles and therefore also vacuum $NS$--tadpoles are
cancelled, as
the tensions of the D-branes and of the orientifold planes precisely balance
each other. 
As we will show this guarantees also the absence of both $R$-- and $NS$--tadpoles 
for one--loop gauge couplings, thus providing $UV$--finite results for these corrections.
On the other hand one--loop gauge corrections in non--supersymmetric intersecting brane
models are plugged by $UV$--divergences. Ultimatively however, one shall be interested in 
non--supersymmetric models. But it seems more convenient, to start with a supersymmetric 
intersecting brane model, in which at least certain couplings are free of $UV$--divergences and 
then breaks supersymmetry by some mechanism, which does not spoil\foot{This procedure is  
possible on the heterotic side, where world--sheet modular invariance guarantees 
$UV$--finiteness even {\it after} supersymmetry breaking. See Ref. \GNS\ for a detailed 
discussion on this problem.} the $UV$--finiteness of the couplings. 
As the main result of this paper we will show that the threshold
corrections in N=1 sectors of intersecting brane world models exhibit a
very interesting, new moduli dependence which is in  contrast to the one-loop threshold
correstions of heterotic N=1 sectors, which are moduli independent.

After a short introduction into the construction of these models 
we work out in section 2 the background gauge dependence of the partition function of the 
open string sectors. 
After a brief review of the technical aspects of orientifolds with supersymmetric 
intersecting branes, we discuss in subsection 3.1. for a given gauge group the possible 
contributions from the various open string diagrams to  its one--loop correction.
These contributions, originating from so--called N=1 and N=2 supersymmetric sectors,
will be calculated in subsections 3.2. and 3.3, respectively.
The calulation for the N=1 sectors need some excursion into the 
mathematical problem of how to extract Dirichlet series from modular forms. 
This allows us to perform certain Schwinger type integrals over Eisenstein series,
presented in appendix \appA. In addition we shall need
some useful spin--structure sums of the gauged open the string partition function, presented
in appendix \appB.
In subsection 3.4 we prove the finiteness of our one--loop gauge threshold results.
This needs the UV--limits of the gauged open string partition functions, which we
exhibit in appendix \appC. We shall see, that in the case of non--anomalous gauge groups 
only NS--tadpoles have to be cancelled. 
The explicit moduli dependence of our N=1 threshold result, given through the radii $R_1^i,R_2^i$ 
rather than the angles of the branes, is shown in subsection 3.6.
In subsection 3.7 we apply our results from subsections 3.2 and 
3.3 to write down the one--loop gauge corrections for a $\IZ_2\times \IZ_2$ orientifold example
with stacks of intersecting branes.

\goodbreak
\newsec{Intersecting branes and gauged open string partition functions}

In subsection 2.1 we review general aspects of intersecting branes 
in IIA toroidal compactifications, which take over to orientifold constructions
without or with orbifold twists.
Some important technical details of the latter, which represent more general brane constructions, 
in particular with the possibility of having N=1 space--time supersymmetry, will be 
reviewed in subsection 2.2.
The main result of this section are the partition functions for open strings stretched
between intersecting branes in the presence of a (space--time) magnetic field. They
will be derived in subsection 2.3. These formulae are quite general and may be used for 
intersecting branes both in toroidal and orientifold/orbifold compactifications of 
type II string theory with arbitrary amount of supersymmetry.

\subsec{Toroidal compactification of intersecting branes}

We consider configurations of type II $D6$ branes wrapped on non--trivial 
three--cycles of a six--dimensional torus $T^6$. 
The gauge theories on the four non--compact dimensions of the brane world--volume
are generically non--chiral and non--supersymmetric, whereas the 
bulk, where all closed string modes live, preserves all 32 supersymmetries of type II.
However placing intersecting branes into an orientifold background (\cf Section 3 for more details)
allows for both non--chiral and chiral  supersymmetric gauge theories.
The torus is taken to be a direct product $T_6=\prod\limits_{j=1}^3 T_2^j$
of three two--dimensional tori $T_2^j$ with radii $R_1^j,R_2^j$ and angles 
$\alpha^j$ w.r.t. to the compact dimensions $e_1^j,e_2^j$. 
The K\"ahler and complex structure modulus of
these tori are defined as usual: 
\eqn\torusmoduli{
U^j=\fc{R_2^j}{R_1^j}e^{i\alpha^j}\ \ ,\ \ T^j=b^j+iR_1^jR_2^j\sin\alpha^j\ ,}
with the torus $B$--field $b^j$.
Furthermore, the three--cycle is assumed to be a factorizable into a direct product of 
three one--cycles, each of them wound around a torus $T_2^j$ with the wrapping
numbers $(n^j,m^j)$ w.r.t. the fundamental 1--cycles of the torus.
Hence the angle of the $D6$--brane with the $Y_1^j$--axis is given by
\eqn\angle{
\tan \phi^j=\fc{m^jR^j_2}{n^jR^j_1}\ .}
Generally, two branes with wrapping numbers $(n^j_a,m^j_a)$ and
$(n^j_b,m^j_b)$, are parallel in the subspace $T_2^j$, 
if their intersection number
\eqn\intersection{
I_{ab}^{j}=\det\pmatrix{n^j_a & m_a^j \cr
                            n^j_b & m_b^j}}
w.r.t. to this subspace vanishes, $I_{ab}^{j}=0$. 
For later convenience let us also introduce:
\eqn\introduce{
\pi v^j:=\ath(\Fc^j)\ ,}
which implies $\phi^j=i\pi\ v^j$.
Chiral fermions appear at (non--vanishing) intersections of two $D6$--branes.

In the $T$--dual picture, the $D6$--branes at angles $\phi^j$ are mapped
to $D9$--branes with magnetic fluxes or background gauge fields $F^j$. 
Thereby the gauge field (magnetic flux) $F^j$ on the brane is related 
to the angles \angle\ through:
\eqn\tdual{
\fc{m^j}{n^j\ R_1^jR_2^j}=:iF^j\ ,}
which results from \angle\ by a $T$--duality in all $Y_2^j$--directions.

\subsec{Some technical aspects of intersecting branes in orbifold/orientifold backgrounds}

The condition for tadpole cancellations in $IIA$ orientifold backgrounds in four space--time 
dimensions requires a system of $D6$ branes which has to respect the orbifold and orientifold 
projections.
In particular this means, that in addition to each brane a system of mirror pairs have 
to be introduced. 
In the models, discussed in \doubref\berlini\bonni, the $D6$ branes are located along 
{\it each} orientifold plane and tadpoles are cancelled locally. 
This leads to a non--chiral N=1 spectrum in $D=4$. However, as realized in \doubref\BGKL\BKLO, this
setup may be relaxed by placing the $D6$ branes at angles w.r.t. only one orientifold plane.
In addition, for consistency with the orbifold/orientifold group their orbifold/orientifold 
mirrors have to be introduced. 
In doing so, it is possible to cancell (non--locally) tadpoles from several orientifold planes 
with fewer branes than if one placed branes on top of each orientifold plane.
The requirement of $R$--tadpole cancellation leads to some constraints on the number and location
of the $D6$ branes. Further restrictions arise in the case of space--time supersymmetry,
where $NS$ vacuum tadpole cancellation follows from $R$ vacuum tadpole cancellation.
One main advantage of allowing for branes with non--vanishing angle
w.r.t. to the orientifold plane is the possibilty of a  chiral N=1 spectrum.

Intersecting brane world models with N=1 supersymmetry in $D=4$
have been introduced in \fourref\berlini\bonni\cvetic\BGO\ through orbifold/orientifold 
projection. The orientifold action $\Om\Rc$ in theses models is a combination
of reversal of world--sheet parity $\Om$ and a reflection of the three internal
coordinates: 
\eqn\reflection{
\Rc:\ \ Y^{2,4,6}\ra -Y^{2,4,6}\ .} 
The orbifold group is generated by elements from $\IZ_N$ (or $\IZ_N\times \IZ_M$). The latter 
are represented by the  $\th$ (and $\om$), describing discrete rotations on the 
compact coordinates $Y^i$. This action restricts the compactification lattice
and fixes some of the internal parameter \torusmoduli\ to discrete values.
The orientifold $O6$--planes describe the set of points which are invariant under the group 
actions $\Om\Rc,\ \Om\Rc\th^k,\ \Om\Rc\om^l$ and $\Om\Rc\th^k\om^l$. These planes are generated
by rotations of the real $Y^{2i-1}$ axes by $\th^{-k/2}\om^{-l/2}$ and will be denoted
by $O6_{k;l}$ in the following.

The models we shall discuss consist of several stacks $a$ of $D6$--branes.  
To each stack, consisting of $N_a$ parallel $D6$ branes, a specific gauge 
group\foot{Before the orientifold and orbifold projection the  gauge group is $U(N_a)$ 
in the case, when the branes are not parallel to an orientifold plane within some subspace. 
Otherwise the gauge group is $USp(N_a)$ or $SO(N_a)$.} $G_a$ is associated.
In addition, in orbifold/orientifold backgrounds, there must exist to each stack $a$ 
a set of orbifold mirrors $\th^n a$ and orientifold $\Om\Rc$--mirrors 
thereof in order to be consistent with the orbifold and orientifold group.
This way we obtain an array of $D$--branes at angles, if the stacks go through a fixpoint of 
the orbifold group, the case we shall consider here.
Hence, following the terminology of \BKLO, any stack $a$ is 
organized in orbits, which represent an equivalence class $[a]$. For $N,M\neq 2$ the length 
of each orbit $[a]$ is at most $2NM$, but may be smaller, if \eg stack $a$ is located along an 
orientifold plane. In that case the length of the orbit is only $MN$.
Since a $\IZ_2$ orbifold twist leaves a brane invariant, in the case of
$N,M=2$, the class $[a]$ consists of just two elements.
Stacks within a conjugacy class $[a]$ have non--trivial intersections among each other and 
w.r.t. to stacks from a different class $[b]$ belonging to the gauge group $G_b$.

The orientifold and orbifold group lead to various open string sectors describing
strings starting and ending on the $D$--branes.
The way, how all these sectors contribute to the vacuum partition function is highly determined
by the requirement of vacuum tadpole cancellation.
In addition some conditions on the representation of the orientifold group
on the Chan-Paton indices follow. 
Open strings starting on a brane from $a$ and ending on one of its orbifolded mirrors 
$\th^n a$ belong to the so--called open string $\th^{2n}$--twisted 
sector. In addition, as known from closed string untwisted orbifold sectors, there exist twist 
insertions $\th^k$ in the partition function, which restricts the contributions to the spectrum
to  $\th^k$ invariant states. This means, that in an annulus diagram $A^{k}_{aa'}$,
which describes an $\th^{2n}$--twisted open string starting on stack $a$ and ending on 
its mirror $a'=\th^n a$, the two stacks $a$ and $a'$ and the orientation of the 
open string have to be invariant under $\th^k$.
In addition we have open string exchanges $A_{ab}^k$ between branes from different classes 
$[a]$ and $[b]$, supplemented with the twist insertion $\th^k$.
Its is evident, that for branes sitting not at orientifold planes a $\th^k$ twist insertion
is only possible, if $\th^k$ represents a $\IZ_2$ twist, \ie for $\th^{N/2},\om^{M/2}$ 
in the case of even order
orbifolds. However, in \doubref\berlini\bonni\ it has been shown, that this statement holds
also for branes sitting at orientifold planes. 
Generically open string annulus diagrams with $\IZ_2$ twist insertions lead
to twisted sector tadpoles in the closed string channel. The latter describing an exchange of a
twisted $5$--form field cannot be cancelled by other amplitudes. 
Therefore one imposes a twisted sector tadpole cancellation condition on the
$\gamma^a_{\th^k}$--matrices acting on the Chan--Paton factors of the open string ends 
\doubref\berlini\bonni:
\eqn\twistedtadpoles{
\Tr \gamma^a_{\th^{N/2}}=0\ \ ,\ \ \Tr \gamma^a_{\om^{M/2}}=0\ \ ,\ \ 
\Tr \gamma^a_{\th^{M/2}\om^{M/2}}=0\ .}
This conditions ensure, that we have not to further discuss those sectors in
the vacuum partition function, and we drop the subscript $k$ on $A_{aa'}^k,A_{ab}^k$.
We shall show in section 3, that this statements takes over for the calculation of 
the one--loop gauge couplings.
The complete annulus partition function takes the form
\eqn\Ages{
A=\sum_{a,b=1}^K
N_aN_b\ \sum_{a \in [a]\atop b\in [b],a\neq b}A_{ab}
=\sum_{a=1,\ldots, K}N_a\ \sum_{b=1\atop b\neq a}^KN_b\ 
\sum_{a \in [a]\atop b\in [b]}A_{ab}+
\sum_{a=1}^KN_a^2\ \sum_{a,a'\in [a]\atop a\neq a'}A_{aa'}\ ,}
with the model dependent numbers $a_{ab}$, which may looked up from 
Refs. \threeref\berlini\bonni\cvetic. The numbers $N_a,N_b$ arise from the traces over
the $\gamma$--matrices acting on the Chan--Paton factors ($\gamma_1^a={\bf 1}_{N_a}$):
\eqn\traces{
\Tr\gamma_1^a=N_a\ .}
We divided the sums into contributions from open string exchanges within one
conjugacy class $[a]$ and exchanges between stacks from different conjugacy classes $[a]$ and 
$[b]$.
In the next subsection we shall see, that this split turns out to be important to 
disentangle the various contributions to the one--loop gauge corrections.

Contributions to the M\"obius partition function may come from untwisted and twisted
sectors with insertion $\Om\Rc\th^k$. In both cases the whole arrangement has to be invariant 
under the action of $\Om\Rc\th^k$.
An untwisted M\"obius  diagram $M^k_{a,a}$ with insertion $\Om\Rc\th^k$ describes a string 
starting and ending on stack $a$. Because of the insertion $\Om\Rc\th^k$, this
exchange may be only possible, if stack $a$ sits on the orientifold plane $O6_{\Om\Rc\th^k}:=O6_k$ 
(see the examples in Ref. \doubref\berlini\bonni).
A M\"obius diagram accounting for a string starting from stack $a$
and ending on its orientifold image $\Om\Rc a$ admits only the twist insertions
$1$ or $\th^{N/2}$, provided stack $a$ does not sit on an orientifold plane. 
We shall denote these two diagrams by $M^0_{a,\Om\Rc a}$ and $M^{N/2}_{a,\th^{N/2}\Om\Rc a}$, 
respectively.
Restrictions on the allowed twist insertion $\th^k$ follow also for twisted M\"obius diagrams.
Namely, only the combinations $M^k_{a,\Om\Rc\th^k a}$ and 
$M^k_{\Om\Rc a,\th^{N-k}a}$ represent twist invariant open string exchanges.
The brane $\Om\Rc\th^k a$ arises from a reflection of brane $a$ on the orientifold plane $O6_k$.
As we shall see in subsection 2.4 these combinations lead to only untwisted sectors 
after transforming into the closed string channel. This is in agreement, that only untwisted 
closed strings interact with the crosscaps.
With $\Om\Rc\th^k=\th^{N-k}\Om\Rc$, we may express all possible M\"obius 
contributions\foot{To keep the formulae readable we only display the 
M\"obius sector for $\IZ_N$ orbifolds. The changes to be made for $\IZ_N\times\IZ_M$ orbifolds 
are straightforward.}  to the vacuum partition function by the following sum
\eqn\Mges{
M=\sum_{a=1}^KN_a\ \sum_{a\in [a]}\ \sum_{k=0}^{N-1} \rho_{\Om\Rc\th^k}\ 
M^k_{a,\Om\Rc\th^k a}\ ,}
with the phases $\rho_{\Om\Rc\th^k}$ as a result from taking the trace
\eqn\tracess{
\Tr[(\gamma_{\Om\Rc\th^k}^{\Om\Rc\th^k a})^\ast\gamma_{\Om\Rc\th^k}^a]=\rho_{\Om\Rc\th^k}\ N_a} 
over the matrices $\gamma_{\Om\Rc\th^k}^a,\gamma_{\Om\Rc\th^k}^{\Om\Rc\th^k a}$
representing the twist action $\Om\Rc\th^k$ on 
brane $a$ and $\Om\Rc\th^ka$, respectively. 
The latter relation generally holds for the Chan--Paton matrices in 
orbifold/orientifold compactifications thanks to the relation:
$(\gamma_{\Om\Rc\th^k}^{\Om\Rc\th^k a})^\ast\gamma_{\Om\Rc\th^k}^a=\rho_{\Om\Rc\th^k}\ 
\gamma^a_{1}=\rho_{\Om\Rc\th^k}\ {\bf 1}_{N_a}$.

\subsec{Gauging the open string partition function} 

All branes we have described above have in common their four--dimensional (non--compact) 
Minkowski space. Hence their gauge fields are located on parallel 
four--dimensional subspaces, which may be seperated 
(in the cases $I_{ab}^j\neq 0$ and $I_{aa'}^j\neq 0$) in the transverse internal 
dimensions. One--loop corrections to the gauge couplings are realized through
exchanges of open strings in that
transverse space. The open string charges $q_a,q_b$ at their ends couple 
to the external gauge fields sitting on the branes.
Only annulus $A$ and M\"obius $M$ diagrams contribute, as torus and 
Klein bottle diagrams refer to closed string states.

When we consider the one--loop correction to the gauge group $G_a$,  at least one open string end
has always to be charged under the  gauge group $G_a$, \ie at least one open string end 
must always end on a brane from stack $a$ (or from its mirrors $\th^n a, \Om\Rc\th^k a,\ldots$).
Open string exchanges between parallel branes preserve N=4 
supersymmetry in the case of no twist insertion.
Therefore, neither $A_{aa}$ nor $M_{aa}^0$ give rise to gauge coupling renormalization. 

Three different kinds of open string exchanges are possible:
An (annulus) exchange $A_{ab}$ between stack $a \in [a]$ constituing to the gauge group 
$G_a$ and stack $b \in [b]$ belonging to the gauge group $G_b$. 
Second, we have to consider (annulus) open string exchanges $A_{aa'}$
between two stacks $a$ and $a'$ from the conjugacy class $[a]$ 
associated to the same gauge group $G_a$. Since $a\neq a'$ this exchange generically
belongs to the open string twisted sector.
Third, there is the whole set of M\"obius diagrams referring to stack $a$ and showing up 
in the sum \Mges\ for $a$.

After having presented formally the various kinds of open string diagrams relevant
for one--loop gauge couplings  in orbifold/orientifold backgrounds with intersecting branes, 
we shall now discuss their gauge background dependent partition functions.
We shall compute the one--loop corrections to the gauge couplings by the background field
method: We turn on a (space--time) magnetic field, \eg $F_{23}=B Q_a$ in the $X^1$--direction 
and determine the dependence of the open string partition function on that field.
Here, $Q_a$ is an appropriatley normalized generator of the gauge group $G_a$ 
under consideration.
The second order of an expansion w.r.t. to $B$ of the gauged partition function  gives 
the relevant piece for the one--loop gauge couplings. 
This procedure has been previously already applied  in \doubref\BF\ABD\ to obtain
one--loop gauge threshold corrections in certain type $I$ orientifold compactifications.
The presence of an external (space--time) gauge field strength 
$F:=F_{23}=BQ_a$, which couples to the charges\foot{More precisely, $q_a,q_{a'}$ are the 
eigenvalues of the group generator $Q_a$ acting on the Chan--Paton states of both string ends.}
$q_a,q_{a'}$ of the open string ends,
results in an shift of the open string oscillator modes by the amount \callan:
\eqn\netshift{
\pi\eps=\arctan(\pi q_a B)+\arctan(\pi q_{a'}B)\ .}
This shift modifies the even\foot{The modification in the odd part of the partition function 
is more involved due to the presence of fermionic zero modes. Eventually we are interested
in $CP$ even gauge couplings. Hence we do not have to worry about gauging odd fermions.} 
spin--structure part of the open string space--time partition function in the following 
way \BP
\eqn\spacepart{\eqalign{
\fc{1}{\eta(\tau)^2}
\fc{ \theta\lf[\de_1\atop\de_2\ri]\lf(0,\tau\ri)}{\eta\lf(\tau\ri)}
&\longrightarrow it\ \be\  
\fc{ \theta\lf[\de_1\atop\de_2\ri]
\lf(\fc{i\eps t}{2},\tau\ri)}{\theta\lf[1/2 \atop  1/2\ri]
\lf(\fc{i\eps t}{2},\tau\ri)}\cr
&=\cases{it\ \be\ e^{-2\pi i\eps(\de_2-\h)}\ 
\fc{\TH{\de_1+\eps}{\de_2}\lf(0,\tau\ri)}
   {\TH{\h +\eps}{\h}
\lf(0,\tau\ri)}\ &,\ \ $\tau=\tau_A=\fc{it}{2}$\ ,\cr
it\ \be\ e^{-2\pi i\eps(\de_2-\h)}\ 
\fc{\TH{\de_1+\eps}{\de_2-\fc{\eps}{2}}\lf(0,\tau\ri)}
   {\TH{\h +\eps}{\h -\fc{\eps}{2}}\lf(0,\tau\ri)}\ 
   &,\ \ $\tau=\tau_M=\fc{it}{2}+\h$\ ,} }}
with:
\eqn\betas{
\be=\pi B\ (q_a+q_{a'})\ .}
Here, $\tau_A$ and $\tau_M$ are the  modular parameter of the annulus and M\"obius
open string partition functions, respectively. 
In the M\"obius amplitude $M_a$ the two endpoint charges are the same, while for
the annulus amplitude $A_{ab}$ describing an open string exchange between one stack
of gauge group $G_a$ an an other stack of different gauge group $G_b$ the second charge
is zero (under the $B$--field under consideration). Thus we have for the net oscillator shifts
\netshift:
\eqn\atn{\eqalign{
A_{ab}:\ \ \     &\pi\eps_{ab}=\arctan(\pi q_a B)\ ,\cr
A_{aa'}:\ \ \    &\pi\eps_{aa'}=\arctan(\pi q_a B)+\arctan(\pi q_{a'} B)\ ,\cr
M_a:\ \ \          &\pi\eps_a=2\arctan(\pi q_a B)\ .}}
In addition we obtain for \betas:
\eqn\charge{\eqalign{
A_{ab}:\ \ \     &\beta_{ab}=\pi B\ q_a\ , \cr
A_{aa'}:\ \ \    &\beta_{aa'}=\pi B\ (q_a+q_{a'})\ ,\cr
M_a:\ \ \          &\beta_a=2\pi B\ q_a\ .}}

Now we have collected all details to gauge the partition function
of an open string stretched between two intersecting stacks
$a$ and $b$ ($a'$, respectively) with rotation angles 
$\phi_a^j$ and $\phi_b^j$ ($\phi_{a'}^j$, respectively) w.r.t. the three complex 
tori $T_2^{j}$. 
The latter have been presented in \BGKL, however may be easily obtained by
applying the procedure \spacepart\ also on the internal open string coordinates:

The gauged open string partition function $A^d_{ab}$ for an open string stretched between 
a stack $a$ associated to the gauge group $G_a$ under consideration and a stack
$b$ of different gauge group becomes:
\eqn\gaugedann{\eqalign{
A^d_{ab}(B)&=i^{d+1}\ \be_{ab}\ t^{d+d'-4}\ \prod\limits_{i=1}^{d'} Z_i(t,T^i,V^i)\cr
&\times\sum_{\vec\de} s_{\vec\de}\ \fc{ \theta\lf[\de_1\atop\de_2\ri]
\lf(\fc{i\eps_{ab} t}{2},\fc{it}{2}\ri)}{\theta\lf[1/2 \atop  1/2\ri]
\lf(\fc{i\eps_{ab} t}{2},\fc{it}{2}\ri)}
\fc{\TH{\de_1}{\de_2}\aarg^{3-d} }{\eta\lf(\fc{it}{2}\ri)^{9-3d}  }\ \prod\limits_{j=1}^d
I^j_{ab}\ \fc{\theta\lf[\de_1\atop\de_2\ri]
\lf(\fc{v_{ab}^j t}{2},\fc{it}{2}\ri)}{\theta\lf[1/2 \atop  1/2\ri]
\lf(\fc{v_{ab}^j t}{2},\fc{it}{2}\ri)}\ .}}
The spin structure sum runs over even spin--structures $\vec\de=(\de_1,\de_2)$, only.
Their phases are given by $s_{(0,0)}=1,\ s_{(0,\h)}=s_{(\h,0)}=-1$.
The quantity $v^j_{ab}$ (\cf \introduce) decribes the relative angle of brane $a$ and $b$
\eqn\angles{
v^j_{ab}:=v_b^j-v_a^j=\fc{i}{\pi}(\phi^j_a-\phi^j_b)}
in close analogy to what happens in the space--time in the presence of the background
magnetic field $B$ (\cf \netshift).
Furthermore, the supscript $d=0,\ldots,3$ denotes the number of internal tori, in which
the brane $a$ and $b$ have non--vanishing intersection $I_{ab}^j\neq 0$,
\ie $v^j_{ab}\neq 0,\pm i$. In the remaining
$d'$ complex internal coordinates they have vanishing intersections $v^i_{ab}= 0,\pm i$. 
Hence w.r.t. these tori $T_2^i$ the open strings have non--vanishing Kaluza--Klein 
momenta and windings. Their mass is given by the mass formula of open strings 
stretched between two parallel $D1$ branes, which are wrapped around the torus $T_2^i$
with wrapping numbers $(n_a^i,m_a^i)$. Their (zero mode) 
partition function has been determined in \doubref\IRAN\BGKLnc
\eqn\zeromodes{
Z_i(t,T^i,V_a^i)=\sum_{r,s\in \IZ}e^{-\fc{\pi t}{T_2^iV_a^i}|r+T^is|^2}\ ,}
with the moduli \torusmoduli\ and: 
\eqn\newmoduli{
V_a^i=\fc{1}{U_2^i}|n_a^i+U^im_a^i|^2=\lf[\fc{R_1^i}{R_2^i}(n_a^i)^2+\fc{R_2^i}{R_1^i}(m_a^i)^2
+2n_a^im_a^i\cos\al^i\ri]\ \fc{1}{\sin\al^i}\ .}
Here and in the following $n_a^i$ and $m_a^i$ are assumed to be coprime integers in order to avoid
multiple wrappings.
Since we focus on intersecting $D6$ branes with a four--dimensional Minkowskian space
we have $d+d'=3$. For the annulus amplitude $A^d_{aa'}$ describing open strings starting and 
ending on stacks $a$ and $a'$ within the same orbit $[a]$ and thus belonging 
to the gauge group $G_a$ we obtain: 
\eqn\gaugedanni{\eqalign{
A^d_{aa'}(B)&=i^{d+1}\ \be_{aa'}\ B\ t^{d+d'-4}\ 
\prod\limits_{i=1}^{d'} Z_i(t,T^i,V^i)\cr
&\times\sum_{\vec\de} s_{\vec\de}\ \fc{ \theta\lf[\de_1\atop\de_2\ri]
\lf(\fc{i\eps_{aa'} t}{2},\fc{it}{2}\ri)}{\theta\lf[1/2 \atop  1/2\ri]
\lf(\fc{i\eps_{aa'} t}{2},\fc{it}{2}\ri)}
\fc{\TH{\de_1}{\de_2}\aarg^{3-d}}{\eta\lf(\fc{it}{2}\ri)^{9-3d}}
\prod\limits_{j=1}^d I_{aa'}^j\ 
\fc{\TH{\de_1}{\de_2}\lf(\fc{v_{aa'}^j t}{2},\fc{it}{2}\ri)}
{\TH{\h}{\h}\lf(\fc{v_{aa'}^j t}{2},\fc{it}{2}\ri)}\ .}}
As before (\cf \angles), $v_{aa'}^j$ is given by the relative angle of the two branes $a$ and 
$a'$:
\eqn\anglesa{
v_{aa'}^j:=v_{a'}^j-v_a^j=\fc{i}{\pi}(\phi_a^j-\phi_{a'}^j)\ .}
Furthermore, for the gauged M\"obius amplitude $M^{k;d}_{a,\Om\Rc\th^k a}(B)$ describing 
an open string starting on a generic stack $a\in[a]$ of the gauge group $G_a$ and 
ending on its orientifold mirror $\Om\Rc\th^k a$ supplemented with the twist insertion $\th^k$, 
we obtain:
\eqn\gaugedmoeb{\eqalign{
M^{k;d}_{a,\Om\Rc\th^k a}(B)&=-i^{d+1}\ \be_a\ t^{d+d'-4}\ \prod\limits_{i=1}^{d'} 
n_{O6_k}^i\ L_i(t,T^i,V_{O6_k}^i)\ 
\sum_{\vec\de} s_{\vec\de}\ \fc{ \theta\lf[\de_1\atop\de_2\ri]
\lf(\fc{i\eps_a t}{2},\fc{it}{2}+\h\ri)}{\theta\lf[1/2 \atop  1/2\ri]
\lf(\fc{i\eps_a t}{2},\fc{it}{2}+\h\ri) }\cr
&\times\prod\limits_{i=1}^{d'}
\fc{\TH{\de_1}{\de_2}\lf(0,\fc{it}{2}+\h\ri)^{3-d}}{\eta\lf(\fc{it}{2}+\h\ri)^{9-3d} }
\ \prod\limits_{j=1}^d2^{\delta_j}\ \fc{I^{k;j}_{a,\Om\Rc\th^k a}\ 
\TH{\de_1}{\de_2}\lf(v_a^{k;j}t,\fc{it}{2}+\h\ri)}
{\TH{\h}{\h}\lf(v_a^{k;j}t,\fc{it}{2}+\h\ri)}\ .}}
Here, $v_a^{k;j}$ is related to the angle $\phi_a^{k;j}$ between brane $a$ 
and the orientifold plane $O6_k$ through:
\eqn\anglesm{
v_a^{k;j}=-\fc{i}{\pi}\phi_a^{k;j}=-\fc{i}{\pi}(\phi^j_a-\phi^j_{O6_k})\ .}
The twist insertion $\th^k$ is automatically respected by choosing $v_a^{k;j}$ in that way.
Of course, for $k=0$, we just have $\phi_a^{0;j}=\phi_a^j$, with $\phi_a^j$ being the angle of
brane $a$ w.r.t. to the positive axis $Y^{2j-1}$.
The factor $2^{\delta_j}$ is an important correction \berlini, relevant if brane $a$ is 
orthogonal to the orientifold plane $O6_k$ in the torus $T_2^j$, \ie $\delta=1$ for 
$\phi_a^{k;j}=\pm\fc{\pi}{2}$.
The intersection number $I_{a,\Om\Rc\th^k a}^{k;j}$ counts the number of 
$\Om\Rc\th^k$--invariant intersections of the two branes $a$ and $a'=\Om\Rc\th^k a$
\eqn\intermoeb{
I^{k;j}_{a;\Om\Rc\th^k a}:=I^j_{a;O6_k}=(n^j_am_{O6_k}^j-m_a^jn^j_{O6_k})\ ,}
with $(n^j_{O6_k},m_{O6_k}^j)$ characterizing the orientifold plane
$O6_k$ w.r.t. to the torus $T_2^j$.
In particular we have \threeref\BlumenhagenEA\cvetic\cveticaa
\eqn\interztwo{
I^{k;j}_{a,\Om\Rc\th^k a}=\cases{-2\ (m_a^j+\h n_a^j)\ ,&$k=0\ ,$\cr  
                                 2^{1-s}\ n_a^j\ ,&$k=1$}}
for a $\IZ_2$ orientifold with ($s=1$) or without ($s=0$) tilted tori $T_2^j$.
In that case the two $O6$--planes in $T_2^j$ are given by $(n^j_{O6_0},m_{O6_0}^j)=(2,0)$ and 
$(n^j_{O6_1},m_{O6_1}^j)=(0,2^{1-s})$.
Contributions $L_i$ from zero modes are possible, 
if brane $a$ and its image  $\Om\Rc \th^k a$ are parallel w.r.t. some torus $i$. 
This is only possible, if brane $a$ sits on the orientifold 
plane $O6_k$ w.r.t. that torus $T_2^i$, \ie $v_a^{k;i}=0$.  
In that case KK momenta parallel and windings orthogonal to the orientifold plane 
$O6_k$ are invariant under the action $\Om\Rc\th^k$ and contribute to 
the sum
\eqn\zeromodesm{
L_i(t,T^i,V_{O6_k}^i)=\sum\limits_{r,s\in \IZ}e^{-\fc{\pi t}{T_2^iV_{O6_k}^i}|r+2^\mu T^is|^2}\ ,}
with $\mu=0$ for $\IZ_2,\IZ_4$ orbifolds with $A$--type lattice and  
$\mu=1$ in the case of $B$--type lattice, and
\eqn\newmoduliO{
V_{O6_k}^i=\lf[\fc{R_1^j}{R_2^j}n_a^jn_{O6_k}^j+\fc{R_2^j}{R_1^j}m_a^jm_{O6_k}^j+
(n_a^jm_{O6_k}^j+n_{O6_k}^jm_a^j)\cos\al^j\ri]\ \fc{1}{\sin\al^j}\ .}
Except for the $\IZ_2$ and $\IZ_4$ orbifold with $A$--type lattice the windings are 
doubled compared to the winding contribution of the annulus diagram \zeromodes.
We refer the reader to the original literature \fourref\berlini\bonni\berlinii\bonnii\ concerning
the possible choices of lattices for a given orbifold action.
In addition, for orbifolds other than $\IZ_2$ we have to choose $R_1^i=R_2^i$.

\subsec{Gauged open string partition functions in the closed string channel}

Finally, in this subsection we present the gauged partition functions, given in {\it Eqs.} 
\gaugedann, \gaugedanni, and \gaugedmoeb\ in the closed string channel, \ie replacing the
modular parameter $t$ by $l=1/t$ and $l=1/(4t)$ for the annulus and M\"obius amplitude, 
respectively. This yields the following expressions, which we shall need in 
the following sections
\eqn\gaugedannc{\eqalign{
\tilde A^d_{ab}(B)&=2^{d-3}(-1)^{d}l\ \be_{ab}\ \prod\limits_{i=1}^{d'} V_a^i\  
\tilde Z_i(l,T^i,V_a^i)\cr
&\times \sum_{\vec\de} s_{\vec\de}\ 
\fc{\TH{\de_1}{\de_2}(\eps_{ab},2il)}{\TH{\h}{\h}(\eps_{ab},2il)}
\fc{\TH{\de_1}{\de_2}\aargc^{3-d}}{\eta(2il)^{9-3d}}\ \prod\limits_{j=1}^d  
I^j_{ab}\ \fc{\TH{\de_1}{\de_2}\lf(iv_{ab}^j,2il\ri)}{\TH{\h}{\h}\lf(iv_{ab}^j,2il\ri)}
\ ,}}
\eqn\gaugedanncc{\eqalign{
\tilde A^d_{aa'}(B)&=2^{d-3}(-1)^{d}l\ \be_{aa'}\ B\ \prod_{i=1}^{d'}V_a^i\ 
\tilde Z_i(l,T^i,V_a^i)\cr
&\times \sum_{\vec\de} s_{\vec\de}\ 
\fc{\TH{\de_1}{\de_2}(\eps_{aa'},2il)}{\TH{\h}{\h}(\eps_{aa'},2il)}
\fc{\TH{\de_1}{\de_2}\aargc^{3-d}}{\eta(2il)^{9-3d}}\ \prod\limits_{j=1}^d I^j_{aa'}\  
\fc{\TH{\de_1}{\de_2}\lf(iv_{aa'}^j,2il\ri)}{\TH{\h}{\h}\lf(iv_{aa'}^j,2il\ri)}\ ,}}
with the resummed lattice function:
\eqn\zeromodesPR{
\tilde Z_i(l,T^i,V_a^i)=\sum_{r,s\in\IZ}e^{-\pi l\fc{V_a^i}{T_2^i}|r+T^is|^2}\ .}
Furthermore for \gaugedmoeb\ we obtain:
\eqn\gaugedmoebc{\eqalign{
\tilde M^{k;d}_{a,\Om\Rc\th^k a}(B)&=-2^{5-d}(-1)^{d}l\ \be_a\ \prod\limits_{i=1}^{d'}
n_{O6_k}^i\ V_{O6_k}^i\ \tilde L_i(4l,T^i,V_{O6_k}^i)\ \sum_{\vec\de} s_{\vec\de}\ 
\fc{\TH{\de_1}{\de_2}\lf(\fc{\eps_a}{2},2il-\h\ri)}
{\TH{\h}{\h}\lf(\fc{\eps_a}{2},2il-\h\ri)}\cr
&\times\prod\limits_{i=1}^{d'}
\fc{\TH{\de_1}{\de_2}\lf(0,2il-\h\ri)^{3-d}}{\eta\lf(2il-\h\ri)^{9-3d}}
\prod\limits_{j=1}^d2^{\delta_j}\ \fc{I_{a,\Om\Rc\th^k a}^{k;j}\ \TH{\de_1}{\de_2}
\lf(iv_a^{k;j},2il-\h\ri)}{\TH{\h}{\h}\lf(iv_a^{k;j},2il-\h\ri)}\ ,}}
with the Poisson resummed expression
\eqn\zeromodesPRm{
\tilde L_i(l,T^i,V_{O6_k}^i)=
\fc{1}{2^\mu}\sum\limits_{r,s\in\IZ}e^{-\pi l\fc{V_{O6_k}^i}{T_2^i}|2^{-\mu} r+T^is|^2}\ ,}
with $\mu=0$ for $\IZ_2,\IZ_4$ orbifolds with $A$--type lattice and 
$\mu=1$ otherwise.

\goodbreak
\newsec{One--loop gauge corrections in supersymmetric intersecting brane world models}

In the following we consider $K$ stacks $a,b,\ldots$ of branes in an N=1 orientifold/orbifold 
background.
Each stack $a$ has $N_a$ parallel branes constituing the gauge group $G_a$
after the orbifold/orientifold group action.
To handle the various mirror images of a given stack under the orientifold/orbifold actions
it proved to be convenient to introduce the notation of a conjugacy class $[a]$ associated 
to each stack $a$. Of course, each member of the conjugacy class $[a]$ appears itself to be 
a stack of $N_a$ copies of the same branes.
(\cf the previous section for further details).

\subsec{Contributions to the one--loop gauge corrections}

To obtain the gauge--threshold corrections w.r.t. a given gauge group $G_a$
one has to extract the order $\Oc(B^2)$ 
part from the gauged partition functions \gaugedann, \gaugedanni\ and \gaugedmoeb. 
A system of intersecting branes may give rise to various sectors with different space--time 
supersymmetries. 
The amount of supercharges which is  preserved by a pair of distinct 
stacks $a$ and $b$ (or two stacks $a,a'$ from the same conjugacy class $[a]$) 
depends on their relative angles $\phi_{ab}^j$ ($\phi_{aa'}^j$) w.r.t. the internal complex 
dimensions, 
given in \eqq \angles. Open string exchanges between parallel branes (\ie $I_{ab}=0$ or 
$I_{aa'}=0$) preserve sixteen 
supercharges and therefore do not lead to any gauge coupling renormalization. 
On the other hand open string exchanges from sectors, which preserve either N=1 or N=2 
supersymmetry, give rise to an non--vanishing one--loop gauge correction. 
Hence, if one $\phi^j_{ab}\neq 0$ (or $\phi^j_{aa'}\neq 0$) in one torus $T_2^j$, 
a portion of the sixteen supersymmetries is broken and we expect non--vanishing 
one--loop gauge couplings for these cases.  

According to the discussion in subsection 2.2 several origins for non--vanishing 
one--loop gauge corrections may be possible:
The first case, where stack $a\in [a]$ associated to the gauge 
group $G_a$ under consideration, preserves N=1 or N=2 supersymmetry with a stack $b\in[b]$ 
from a different stack $b$.
These cases, collectively denoted as $ab$--exchange, are described by the gauged 
annulus partition functions $A_{ab}^{d=3,2}(B)$, presented in \gaugedann.
They give rise to the one--loop gauge corrections $\Delta^{N=1,2}_{ab}$, respectively.
Second, N=1 or N=2 supersymmetric sectors are possible for open string exchanges
between two stacks $a,a'\in [a]$ stemming from the same conjugacy class. These diagrams,
summarized as $aa'$--exchange and described by the amplitude $A_{aa'}^{d=3,2}(B)$, 
lead to the corrections  $\Delta^{N=1,2}_{aa'}$. 
Finally, depending on the amount of supersymmetry preserved by the two stacks $a$ and 
$\Om\Rc\th^ka$, the gauged M\"obius diagram $M_{a,\Om\Rc\th^ka}^k(B)$ gives rise to
the one--loop gauge corrections\foot{In the case of a M\"obius diagram we shall always keep 
the twist order $k$ as subscript on the correction $\Delta^{k;N=1,2}_{a,\Om\Rc\th^ka}$ in 
order to distungish it from the correction from an annulus diagram 
$\Delta^{N=1,2}_{aa'}$.}  
$\Delta^{k;N=1}_{a,\Om\Rc\th^ka}$ and $\Delta^{k;N=2}_{a,\Om\Rc\th^ka}$, respectively.
The latter describe open strings starting on brane $a$ and ending on its orientifold
image $\Om\Rc\th^ka$.

To summarize, let us present the complete one--loop correction to the gauge coupling 
$g_a^{-2}$ of the gauge group $G_a$
\eqn\oneloop{\eqalign{
\Delta_{G_a}&=\sum_{b=1\atop b\neq a}^K\ \sum_{a \in [a]\atop b\in [b]}
\Delta_{ab}+\ \sum_{a,a'\in [a]\atop a\neq a'}\Delta_{aa'}
+\sum_{a\in [a]}\ \sum_{k=0}^{N-1}\ \Delta^k_{a,\Om\Rc\th^k a}\ ,}}
with the Schwinger integrals (converted into the closed string channel):
\eqn\schwinger{\eqalign{
\Delta^{N=1,2}_{ab}&=2\pi^{-2}\int\limits_0^\infty\ \fc{dl}{l}\ \Tr\lf.\fc{\p^2}{\p B^2}\  
\tilde A_{ab}^{d=3,2}(B)\ri|_{B=0}\ ,\cr
\Delta^{N=1,2}_{aa'}&=2\pi^{-2}\int\limits_0^\infty\ \fc{dl}{l}\ \Tr\lf.\fc{\p^2}{\p B^2}\  
\tilde A_{aa'}^{d=3,2}(B)\ri|_{B=0}\ ,\cr
\Delta^{k;N=1,2}_{a,\Om\Rc\th^ka}&=\pi^{-2}
\int\limits_0^\infty\ \fc{dl}{l}\ \Tr\lf.\fc{\p^2}{\p B^2}\  
\tilde M^{k;d=3,2}_{a,\Om\Rc\th^ka}(B)\ri|_{B=0}\ .}}
The factor $2$ in the front of the interals accounts for the two different orientations of
the open strings w.r.t. to stack $a$.
The group trace $\Tr$ is accomplished by summing over all string endpoint charges 
$q_a,q_b$ ($q_{a'}$, respectively) 
augmented with the orientifold projection\foot{The phase $\rho_{\Om\Rc\th^k}$ in the sum \Mges, 
which arises from \tracess, will show up again 
after performing the traces within the intergal $\Delta^{k;N=1,2}_{a,\Om\Rc\th^ka}$.}. 
We will be more precise about that
in the following subsections. In addition, on the symbol $\Delta$ we have put the subscript $N=1,2$
referring to the amount of supersymmetries preserved by the two branes involved. 
As described before, N=1,2 supersymmetries are  respectively related to $d=3,2$ tori 
$T_2^j$, in which the branes have non--vanishing intersections. 
The sum \oneloop\ represents a sum over various open string sectors in close
analogy to the expressions \Ages\ and \Mges. 
However, the sum \oneloop\ includes much fewer sectors, since at least 
one open string end has to couple to a brane of gauge group $G_a$. In addition
N=4 sectors do not contribute.

After these preparations let us now determine the general form of the one--loop
gauge threshold correction \oneloop\ by evaluating the Schwinger integrals \schwinger\ 
for N=1 and N=2 supersymmetric sectors of intersecting brane world models, leading to the
results for  $\Delta^{\rm N=1,2}_{ab},\Delta^{\rm N=1,2}_{aa'}$ and 
$\Delta^{k,\rm N=1,2}_{a,\Om\Rc\th^ka}$, respectively.
This will be accomplished in several steps: From the 
closed string expressions \gaugedannc, \gaugedanncc, and \gaugedmoebc\  
entering \schwinger\ we determine that piece which is second order in the magnetic field $B$, 
perform the spin--structure sum and eventually do the integration.
In additon we must clarify the possible existence of \UV divergences in \oneloop.
We give the general rules for vanishing $UV$--divergences  in subsection 3.4.
In subsection 3.7 we apply our results to present the gauge thresholds for a 
$\IZ_2\times \IZ_2$ orientifold model.

\subsec{Gauge thresholds from N=1 supersymmetric sectors}

Let us first determine $\Delta_{ab}^{\rm N=1}$ for the case when stack $a$ and $b$ 
preserve N=1 supersymmetry. 
To extract from $\tilde A_{ab}(B)$ the second order in the magnetic field $B$ we use
\eqn\expand{\eqalign{
\pi\ q_a&\lf.\fc{\p^2}{\p B^2}\  B\ \fc{\th\lf[\de_1\atop\de_2\ri](\eps_{ab},\tau)}
{\th\lf[\h\atop\h\ri](\eps_{ab},\tau)}\ri|_{B=0}\cr
&=-\fc{\pi^2\ q_a^2}{\eta^3}\lf\{\fc{1}{3}\th\lf[\de_1\atop\de_2\ri](0,\tau)+\fc{1}{6}
E_2(\tau) \th\lf[\de_1\atop\de_2\ri](0,\tau)+\fc{1}{2\pi^2}
\th''\lf[\de_1\atop\de_2\ri](0,\tau)\ri\},}}
with $e^{2\pi i \eps_{ab}}=\fc{1+i\pi q_a B}{1-i\pi q_a B}$, given in \eqq \atn. 
For the N=1 supersymmetric sector only the last term of \expand\ gives
rise to a non-vanishing contribution to \oneloop\ after performing the spin--structure
sum. That contribution we have calculated in the \eqq (B.2). Hence  
the gauge threshold corrections for an N=1 sector take\foot{
In the following, from the possible realizations of N=1 supersymmetry, 
$\pm v^1_{ab}\pm v^2_{ab}\pm v^3_{ab}=0$, we investigate the case 
$v^1_{ab}+v^2_{ab}+v^3_{ab}=0$.
The other cases follow from the latter by changing signs.} the form
\eqn\ti{
\Delta^{\rm N=1}_{ab}=-2\pi^{-1}\ b_{ab}^{N=1}\ \int_0^\infty  dl\ \lf[
\fc{\th_1'(iv^1_{ab},2il)}{\th_1(iv^1_{ab},2il)}+
\fc{\th_1'(iv^2_{ab},2il)}{\th_1(iv^2_{ab},2il)}+
\fc{\th_1'(iv^3_{ab},2il)}{\th_1(iv^3_{ab},2il)}\ri]\ ,}
for the case $v^1_{ab}+v^2_{ab}+v^3_{ab}=0$. 
Here the N=1
$\beta$--function coefficient is given by:
\eqn\BETAa{
b_{ab}^{N=1}=I_{ab}\ \Tr(Q_a^2\gamma_1^a\otimes \gamma^b_1)\ .}
We reinstated the charge operator $Q_a$ of the gauge group $G_a$ and the matrices 
$\gamma^a_1,\gamma^b_1$ acting on the Chan--Paton states at the endpoints of the open strings.
They fulfill $\gamma^b_1={\bf 1}_{N_b}$ \doubref\berlini\bonni, since
$N_b$ branes sit on top of each brane from stack $b$.
The order of $Q_a$ and $\gamma^a_1$ in the trace is not relevant, since $Q_a$ commutes with 
$\gamma^a_1$. 
Otherwise the gauged open string state would have not survived the orientifold projection.
With this information we obtain:
\eqn\BETAA{b_{ab}^{N=1}=I_{ab}\ \Tr(Q_a^2\gamma^a_1)\ \Tr(\gamma_1^b)=
N_b\ I_{ab}\ \Tr(Q_a^2)\ .}
At this point, we easily see why annulus diagrams $A_{ab}$ with $\IZ_2$ twist insertions
do not give rise to one--loop gauge corrections. Due to \twistedtadpoles, \ie 
$\Tr\gamma_{\th^{N/2}}^a=0$, following
from the cancellation of twisted sector tadpole contributions, annulus 
diagrams with $\IZ_2$ twist insertions give rise
to vanishing $\beta$--function coefficients, \ie $\Delta_{ab}=0$.
Finally, we have defined: $I_{ab}=\prod\limits_{j=1}^3I_{ab}^j$.

The integrand of \ti\ does not change when $iv_{ab}^j$ is shifted into the 
region $-\pi <iv^j_{ab}<\pi$ by integers of $\pm 1$. 
Hence we may perform shifts on the angles $\phi_{ba}^j$ such that eventually $|iv_{ab}|<\pi$ 
is achieved. However only those shifts are allowed, which preserve the condition 
$v^1_{ab}+v^2_{ab}+v^3_{ab}=0$.
To evaluate \ti\ we start from the relation
\eqn\nontrivial{
\fc{\th_1'(iv,\tau)}{\th_1(iv,\tau)}=-i\fc{\p}{\p v}
\ln\th_1(iv,\tau)=-\pi i\coth(\pi v)-\sum_{k=1}^\infty \alpha_{2k}\ (iv)^{2k-1}\  
[E_{2k}(\tau)-1]\ ,}
which can be derived from the identities \LSW:
\eqn\nice{
\fc{z\th_1'(0,\tau)}{\th_1(z,\tau)}=
e^{\sum\limits_{k=1}^\infty \fc{\alpha_{2k}}{2k}\ z^{2k}\ E_{2k}(\tau)}\ \ \ ,\ \ \ 
\alpha_{2k}=-\fc{(2\pi i)^{2k}}{(2 k)!} B_{2k}=2\zeta(2k)\ ,}
and:
\eqn\nicei{
\sum_{k=1}^\infty \fc{\alpha_{2k}}{2k}\ z^{2k}=\ln(\pi z)-\ln\sin(\pi z)\ \ ,
\ \ 0<z<1\ .}

After inserting \nontrivial\ into \ti, the first term of \nontrivial\ 
gives rise to the following apparent divergent\foot{$UV$ divergent in the open string 
sector.} integral:
\eqn\followint{
\delta_{ab}^{\rm N=1}=2i\ b_{ab}^{N=1}\ \lf[\coth(\pi v^1_{ab})+\coth(\pi v^2_{ab})+
\coth(\pi v^3_{ab})\ri]\ \int_0^\infty\ dl\ ,}
with (\cf \eqq (C.3)):
\eqn\Relation{
\coth(\pi v_{ab}^j)=i\cot(\phi_b^j-\phi_a^j)=
i\ \fc{n_a^jn_b^j\fc{R_1^j}{R_2^j}+m_a^jm_b^j\fc{R_2^j}{R_1^j}+(n_a^jm_b^j+n_b^jm_a^j)
\cos\alpha^j}{n^j_a m_b^j-n^j_b m^j_a}\ \fc{1}{\sin\al^j}\ .}
The expression \followint\ can be identified with the $UV$ divergence stemming 
from the $NS$ sector, derived in \eqq (C.11) (\cf subsection 3.4). Note, the additional factor
of $2$ in \followint\ from the two different open string orientations.
It will prove to be important for the cancellation of $NS$ tadpole contributions in the 
complete expression \oneloop, after taking into account similar divergences 
from the other sectors. We shall discuss them later and discard $\delta_{ab}^{\rm N=1}$ 
for a short moment to come back to it in subsection 3.4.

Let us now proceed with the sum of \nontrivial, which represents the pure string contribution
and gives rise to the following type of 
integrals:
\eqn\TYP{
\int\limits_0^\infty  dy\ [E_{2n}(iy)-1]\ .}
A naive integration of the Eisenstein function $E_{2n}(iy)$, given as power series in 
$e^{-2\pi y}$,  would lead to a non--converging series.
The situation is to be contrasted with integrals over the torus 
fundmantal region of modular invariant functions as they arise from closed string 
one--loop amplitudes \LNSW.
In these case, the \UV region $y\ra 0$ is excluded, which allows to use an expansion of
the modular function w.r.t to $e^{-2\pi y}$ and leads to finite results.
On more mathematical grounds, the integral \TYP\ represents an isomorphism
between a modular form and a Dirichlet series. This problem has been studied
by Hecke and we will apply his methods in Appendix A.
In fact, there we shall see, that the integrals \TYP\ can be
evaluated after analytic continuation with the result\foot{We use $\zeta(2n)=
\fc{1}{(2n)!}\ 2^{2n-1}\pi^{2n}|B_{2n}|$.} (A.16):
\eqn\form{\eqalign{
\int\limits_0^\infty  dl\ [E_{2n}(2il)-1]&=
\fc{2^{-2n}\pi^{1-2n}(2n)!}{(1-2n)|B_{2n}|}\ \zeta(2n-1)\ ,\ n>1\ ,\cr
\int\limits_0^\infty  dl\ l^\eps\ [E_2(2il)-1]&=\fc{3}{\pi\eps}-\fc{3}{\pi}\ln 2\ .}}
Essentially the parameter $\eps>0$ originates from a dimensional regularization
(cf. also \FS).
The term $\fc{3}{\pi\eps}-\fc{3}{\pi}\ln 2$ accounts for the infrared divergence of massless 
closed string modes. The summand $k=1$ of \nontrivial\ is linear in $iv$, 
which thanks to supersymmetry, vanishes in the expression \ti\ after summing over all three 
complex planes. Hence we may drop the case $k=1$.
With the results \form\ the last sum of \nontrivial\ can be integrated to:
\eqn\niceintegral{\eqalign{
\int\limits_0^\infty  dl\ \sum_{k=2}^\infty \alpha_{2k}\ (iv)^{2k-1}\ [E_{2k}(2il)-1]
&=-\pi \ \sum_{k=2}^\infty (iv)^{2k-1}\ \fc{\zeta(2k-1)}{2k-1}\cr
&=-\h\pi \ \ln\lf[\ e^{-2i\gamma_Ev} \ \fc{\Gamma(1-iv)}{\Gamma(1+iv)}\ \ri]\ .}}
For the last step we used the identity \Grad:
\eqn\grad{
\sum_{n=1}^\infty\fc{x^{2n+1}}{2n+1}\ \zeta(2n+1)=\h\ln\lf[e^{-2\gamma_E x}
\fc{\Gamma(1-x)}{\Gamma(1+x)}\ri]\ \ ,\ \ x\in \IR\ ,\ |x|<1\ .}
After these preparations we are ready to determine \ti.
For $|iv_{ab}^j|=\fc{1}{\pi}|\phi^j_{ba}|<1$ we obtain:
\eqn\altertiinew{
\Delta^{\rm N=1}_{ab}=\delta_{ab}^{\rm N=1}
-b_{ab}^{N=1}\  \ln \fc{\Gamma(1-\fc{1}{\pi}\phi_{ba}^1)\ 
\Gamma(1-\fc{1}{\pi}\phi_{ba}^2)\ \Gamma(1+\fc{1}{\pi}\phi_{ba}^1+
\fc{1}{\pi}\phi_{ba}^2)}{\Gamma(1+\fc{1}{\pi}\phi_{ba}^1)\ 
\Gamma(1+\fc{1}{\pi}\phi_{ba}^2)\ \Gamma(1-\fc{1}{\pi}\phi_{ba}^1-
\fc{1}{\pi}\phi_{ba}^2)}  \ .}
Here the moduli dependence is given implicitly through the relation \Relation. 
In subsection 3.6 we shall present an alternative expression for \altertiinew\ showing its 
explicit dependence on the radii $R_1^i,R_2^i$. Due to the symmetry behaviour
of $b_{ab}^{N=1}$, following from $I_{ab}\leftrightarrow -I_{ab}$ under the exchange of the 
two branes $a\leftrightarrow b$,
the whole result $\Delta^{\rm N=1}_{ab}$ does not alter under permuting the brane $a$ and $b$.

Let us now come to the M\"obius sector \gaugedmoeb, describing an open string
starting from brane $a$ and ending on $a'=\Om\Rc\th^ka$.
Similar as before, using \eqq \expand\ and the spin--structure sum (B.2), the correction
from the N=1 sector can be expressed by ($\tau=2il-\h$)
\eqn\tiim{
\Delta^{k;\rm N=1}_{a,\Om\Rc\th^k a}=-4\pi^{-1}\ b^{k;N=1}_{a,\Om\Rc\th^k a}\ 
\int_0^\infty  dl\ \lf[
\fc{\th_1'(iv^{k;1}_a,\tau)}{\th_1(iv^{k;1}_a,\tau)}+
\fc{\th_1'(iv^{k;2}_a,\tau)}{\th_1(iv^{k;2}_a,\tau)}+
\fc{\th_1'(iv^{k;3}_a,\tau)}{\th_1(iv^{k;3}_a,\tau)}\ri]\ ,}
for the case $v^{k;1}_a+v^{k;2}_a+v^{k;3}_a=0$, with the $\beta$--function coefficient:
\eqn\BETAm{
b^{k;\rm N=1}_{a,\Om\Rc\th^k a}=-2 I^{k}_{a,\Om\Rc\th^k a}\ 
\Tr[Q_a^2\ (\gamma_{\Om\Rc\th^k}^{\Om\Rc\th^k a})^\ast\gamma_{\Om\Rc\th^k}^a]=
-2\ N_a\ \rho_{\Om\Rc\th^k}\ I_{a;O6_k}\ \Tr(Q_a^2)\ .}
We used \tracess\ to perform the trace over the $\gamma$--matrices.
As before for $\Delta^{\rm N=1}_{ab}$, with the help of \nontrivial, we may disentangle 
$UV$--divergent contribution of \tiim: 
\eqn\followintm{
\delta_{a,\Om\Rc\th^k a}^{k;\rm N=1}=4i\ b^{k;N=1}_a\ 
\lf[\coth(\pi v^{k;1}_a)+\coth(\pi v^{k;2}_a)+\coth(\pi v^{k;3}_a) \ri]\ \int_0^\infty\ dl\ .}
In addition, we have defined $I_{a;O6_k}=\prod\limits_{j=1}^3I_{a;O6_k}^j$.
With the integrals
\eqn\knowledge{\eqalign{
\int\limits_0^\infty  dl\ [E_{2n}(2il-\h)-1]&=
2^{2n-4}\ \fc{2^{-2n}\pi^{1-2n}(2n)!}{(1-2n)|B_{2n}|}\ \zeta(2n-1)\ ,\ n>1\ ,\cr
\int\limits_0^\infty  dl\ l^\eps\ [E_2(2il-\h)-1]&=\fc{3}{4\pi\eps}+\fc{3}{4\pi}\ln 2}}
following from (A.17) the finite part of \tiim\ can be integrated to give a similar expression 
than 
\altertiinew 
\eqn\altertiimm{
\Delta^{k;\rm N=1}_{a,\Om\Rc\th^k a}=\delta_{a,\Om\Rc\th^k a}^{k;\rm N=1}
-\fc{1}{4}\ b^{k;\rm N=1}_{a,\Om\Rc\th^k a}\  \ln \fc{\Gamma(1-\fc{2}{\pi}\phi_a^{k;1})\ 
\Gamma(1-\fc{2}{\pi}\phi_a^{k;2})\ \Gamma(1+\fc{2}{\pi}\phi_a^{k;1}+\fc{2}{\pi}\phi_a^{k;2})}
{\Gamma(1+\fc{2}{\pi}\phi_a^{k;1})\ \Gamma(1+\fc{2}{\pi}\phi_a^{k;2})\ 
\Gamma(1-\fc{2}{\pi}\phi_a^{k;1}-\fc{1}{\pi}\phi_a^{k;2})}  \ ,}
with the hidden moduli dependence encoded in the angles \anglesm\ through through:
\eqn\Relationm{\eqalign{
\coth(\pi v_a^{k;j})&=i\cot(\phi_a^{k;j})=i\cot(\phi_a^j-\phi^j_{O6_k})\cr
&=\fc{i}{I^{j}_{O6_k,a}}\ 
\lf[n_{O6_k}^jn_a^j\fc{R_1^j}{R_2^j}+m_{O6_k}^jm_a^j\fc{R_2^j}{R_1^j}+(n^j_{O6_k}m_a^j+
n_a^jm^j_{O6_k})\cos\al^j\ri]\fc{1}{\sin\al^j}\ .}}
Again, the angles $\phi_a^{k;j}$ have to be shifted into the domain 
$-\fc{\pi}{2}<\phi_a^{k;j}<\fc{\pi}{2}$ by integers of $\pm \pi$. 
However only those shifts are allowed, which do not alter the integrand
of \tiim\ and obey the supersymmetry condition $\phi_a^{k;1}+\phi_a^{k;2}+\phi_a^{k;3}=0$.

Finally, let us move on to the corrections $\Delta_{aa'}$ arsing from an open string 
exchange between a stack $a$ and one of its mirrors $a'\in[a]$. 
We first note the identity
\eqn\expandi{\eqalign{
\pi\ (q_a+q_{a'})&\lf.\fc{\p^2}{\p B^2}\  B\
\fc{\th\lf[\de_1\atop\de_2\ri](\eps_{aa'},\tau)}
{\th\lf[\h\atop\h\ri](\eps_{aa'},\tau)}\ri|_{B=0}\cr
&=-\pi^2\fc{q_a^2+q_{a'}^2}{\eta^3}
\lf\{\fc{1}{3}\th\lf[\de_1\atop\de_2\ri](0,\tau)+\fc{1}{6}
E_2(\tau) \th\lf[\de_1\atop\de_2\ri](0,\tau)+\fc{1}{2\pi^2}
\th''\lf[\de_1\atop\de_2\ri](0,\tau)\ri\}\cr
&+\pi^2\fc{q_a q_{a'}}{\eta^3}
\lf\{\fc{1}{3}\th\lf[\de_1\atop\de_2\ri](0,\tau)-\fc{1}{3}
E_2(\tau) \th\lf[\de_1\atop\de_2\ri](0,\tau)-\fc{1}{\pi^2}
\th''\lf[\de_1\atop\de_2\ri](0,\tau)\ri\}\ ,}}
with $e^{2\pi i \eps_{aa'}}=
\fc{(1+i\pi q_a B)(1+i\pi q_{a'}B)}{(1-i\pi q_a B)(1-i\pi q_{a'}B)}$,
from \eqq \atn.
This relation allows us to extract the $\Oc(B^2)$ part from $\tilde A_{aa'}(B)$. 
We do not discuss any further the second term of \expandi, since its factor in front 
has a charge combination, which corresponds to the trace:
$\Tr(Q_{a}\gamma^a_1)\Tr(Q_{a'}\gamma^{a'}_1)=0$. These traces vanish 
\eqn\anomalyfree{
\Tr(Q_{a}\gamma^a_1)=\Tr(Q_{a})=0}
under the assumption, that the gauge group $G_a$ under consideration is not one of 
the possible anomalous $U(1)$ gauge groups of the theory under consideration.
Putting the first piece of \expandi\ into the spin--structure sum 
(\cf appendix \appBi) yields ($\tau=2il$):
\eqn\threshmirror{\eqalign{
\Delta_{aa'}^{\rm N=1}=-2\pi^{-1}\ b^{N=1}_{aa'}\int_0^\infty  dl\ \lf[
\fc{\th_1'(iv^1_{aa'},\tau)}{\th_1(iv^1_{aa'},\tau)}+
\fc{\th_1'(iv^2_{aa'},\tau)}{\th_1(iv^2_{aa'},\tau)}+
\fc{\th_1'(iv^3_{aa'},\tau)}{\th_1(iv^3_{aa'},\tau)}\ri]\ .}}
with the $\beta$--coefficient
\eqn\betaaa{
b_{aa'}^{N=1}=I_{aa'}\ 
[\Tr(Q_a^2\gamma^a_1)\Tr(\gamma^{a'}_1)+\Tr(\gamma^{a}_1)\Tr(Q_a^2\gamma^{a'}_1)]
=2\ N_a\ I_{aa'}\ \Tr(Q_a^2)\ .}
The integral \threshmirror\ contains an \UV divergence analogous to \followint:
\eqn\followintmirror{
\delta_{aa'}^{\rm N=1}=2i\ b^{N=1}_{aa'}\ 
\lf[\coth(\pi v_{aa'}^1)+\coth(\pi v_{aa'}^2)+\coth(\pi v_{aa'}^3)\ri]\ \int_0^\infty\ dl\ ,}
which entirely steams from the $NS$--sector and will be further discussed in 
subsection 3.4. 
Like in the previous cases the integral \threshmirror\ yields:
\eqn\altertiiaa{
\Delta^{\rm N=1}_{aa'}=\delta_{aa'}^{\rm N=1}
-b^{\rm N=1}_{aa'}\  \ln \fc{\Gamma(1-\fc{1}{\pi}\phi_{a'a}^1)\ 
\Gamma(1-\fc{1}{\pi}\phi_\as^2)\ \Gamma(1+\fc{1}{\pi}\phi_\as^1+\fc{1}{\pi}\phi_\as^2)}
{\Gamma(1+\fc{1}{\pi}\phi_\as^1)\ \Gamma(1+\fc{1}{\pi}\phi_\as^2)\ 
\Gamma(1-\fc{1}{\pi}\phi_\as^1-\fc{1}{\pi}\phi_\as^2)}  \ ,}
In analogy to \Relation\ the difference of the angles $\phi^j_a$ and $\phi^j_{a'}$ are 
related to the radii through:
\eqn\relationaa{
\coth(\pi v_{aa'}^j)=i\cot(\phi_{a'}^j-\phi_a^j)=
i\ \fc{n_a^jn_\ap^j\fc{R_1^j}{R_2^j}+m_a^jm_\ap^j\fc{R_2^j}{R_1^j}+(n_a^jm_\ap^j+n_\ap^jm_a^j)
\cos\alpha^j}{n^j_a m_\ap^j-n^j_\ap m^j_a}\ \fc{1}{\sin\al^j}\ .}

\subsec{Gauge thresholds from N=2 supersymmetric sectors}

The discussion of N=2 sectors is somewhat much simpler as for N=1 sectors since
branes of this sector represent $1/2$ BPS saturated objects and therefore only massless open 
string states contribute to the gauge coupling renormalization. This manifests in drastic 
simplifications  in the gauged open string partition functions for 
this sector (at second order in the magnetic field $B$). 
For N=2 sectors, whose spin--structure sum has been performed in appendix \appBii, 
we have $v_{ab}^i=0$ w.r.t. the $i$--th plane and  $v^j_{ab}\pm v^l_{ab}=0$ for the two 
remaining planes. 
With \eqq (B.3)  the second integral of \schwinger\ boils down to:
\eqn\tii{
\Delta^{\rm N=2}_{ab}=b^{\rm N=2}_{ab}\ V_a^i\ \int_0^\infty  dl\ \tilde 
Z_i(l,T^i,V_a^i)\ ,}
with the N=2 $\beta$--function coefficient:
\eqn\betaii{
b^{\rm N=2}_{ab}=-2I^j_{ab}I^l_{ab}\ \Tr(Q_a^2\gamma^a_1)\Tr(\gamma^b_1)=
-2N_b\ I^j_{ab}I^l_{ab}\  \Tr(Q_a^2)\ .}
Again as before the same argument about possible $\IZ_2$ twist insertions applies:
With \twistedtadpoles\ for such sectors the $\beta$--function coefficient and hence
also $\Delta^{\rm N=2}_{ab}$ vanish. A divergence 
\eqn\followintii{
\delta_{ab}^{\rm N=2}=b^{\rm N=2}_{ab}\ V_a^i\int_0^\infty dl}
due to the zero momentum state $(r,s)=(0,0)$ in \zeromodesPR\ is encountered in \tii\ for the 
limit $l\ra\infty$.
After inspection of (C.10) it is identified as a potential tadpole contribution in 
the $NS$--sector.
As before for $\delta_{ab}^{\rm N=1}$, we shall discuss its relevance in subsection 3.4,
where we shall conclude that in the complete expression \oneloop\ all $UV$--divergences are 
cancelled.
Hence we split that term from the integral $\Delta^{\rm N=2}_{ab}$ and write
\eqn\Tii{\eqalign{
\Delta_{ab}^{\rm N=2}&=\delta_{ab}^{\rm N=2}+b^{\rm N=2}_{ab}\ V_a^i\ 
\int_0^\infty  dl\ 
\sum_{(r,s)\neq (0,0)}e^{-\pi l\fc{V^i_a}{T_2^i}|r+T^ks|^2}\cr
&=\delta_{ab}^{\rm N=2}-b_{ab}^{N=2}\ \lf[\ln T^i_2|\eta(T^i)|^4+\ln V_a^i-
\kappa\ri]\ ,}}
with the constant $\kappa=\gamma_E-\ln(4\pi)$. The last integral has been already performed
in Refs. \DKL. Its $IR$--regularization is achieved by some sort of dimensional
regularization \FS. 
We conclude, that apart from its second term $\ln V_a^i$ the functional form of the gauge 
threshold correction $\Delta_{ab}^{\rm N=2}$ is the same as we know already from  N=2 type 
$I$ orientifold compactifications \doubref\BF\ABD. 

Not much changes for the open string exchange between stack $a$ and one of its mirrors $a'$.
For this case we obtain the one--loop gauge correction
\eqn\aaprime{\eqalign{
\Delta^{\rm N=2}_{aa'}&=b^{\rm N=2}_{aa'}\ V_a^i\ \int_0^\infty  dl\ \tilde Z_i(l,T^i,V_a^i)\cr
&=\delta_{aa'}^{\rm N=2}-b^{N=2}_{aa'}\ 
\lf[\ln T^i_2|\eta(T^i)|^4+\ln V_a^i-\kappa\ri]\ ,}}
with the N=2 $\beta$--function coefficient (\cf the arguments leading to \eqq \betaaa):
\eqn\betamiia{
b^{\rm N=2}_{aa'}=
-2I^j_{aa'}I^l_{aa'}\ [\Tr(Q_a^2\gamma^a_1)\Tr(\gamma^{a'}_1)+
\Tr(\gamma^{a}_1)\Tr(Q_a^2\gamma^{a'}_1)]=-4\ N_a\ I^j_{aa'}I^l_{aa'}\ \Tr(Q_a^2)\ .}
Again, in the $NS$ sector we have to face a divergence 
\eqn\divaa{
\delta_{aa'}^{N=2}=b^{\rm N=2}_{aa'}\ V_a^i\int_0^\infty dl} 
from the zero momentum state $(r,s)=(0,0)$, to comment on later.

Finally let us come to the M\"obius sector $M^k_{a;\Om\Rc\th^k a}$ with N=2 supersymmetry, \ie
brane $a$ and its image $\Om\Rc\th^k a$ are parallel within one torus $T_2^i$, 
\ie $\phi_a^{k;i}=0$. The insertion $\Om\Rc\th^k_i$ leaves invariant momenta parallel and windings
orthogonal to the orientifold plane $O6_k$ w.r.t. to the $i$--th plane. These states are 
encoded in the lattice sum $L_i(t,T^i,V_{O6_k}^i)$, given in \zeromodesm.
In that case the relevant gauged partition function is \gaugedmoebc, with $d'=1, d=2$. Its
second order in $B$ may be obtained from \eqq (B.3). 
It gives rise to the one--loop gauge correction 
\eqn\tmoeb{\eqalign{
\Delta^{k;\rm N=2}_{a;\Om\Rc\th^k a}&=4\ b^{k;N=2}_{aa'}\ V_{O6_k}^i\ \int_0^\infty  dl\ 
\tilde L_i(4l,T^i,V_{O6_k}^i)\cr
&=\delta_{aa'}^{k;\rm N=2}-b^{k;N=2}_{aa'}\ 
\lf[\ln T^i_2|\eta(2^{\mu}T^i)|^4+\ln V_{O6_k}^i+\ln 4-\kappa\ri]\ ,}}
with the N=2 $\beta$--function coefficient (\cf \tracess) :
\eqn\betamii{\eqalign{
b^{k;N=2}_{aa'}&=8\ I^j_{O6_k;a}I^l_{O6_k;a}\ 
\Tr[Q_a^2\ (\gamma_{\Om\Rc}^{\Om\Rc a})^\ast\gamma_{\Om\Rc}^a]\cr
&=8\ N_a\ \rho_{\Om\Rc\th^k}\ I^j_{O6_k;a}\ I^l_{O6_k;a}\ \Tr(Q_a^2)\ .}}
The potential tadpole contribution
\eqn\divm{
\delta_{aa'}^{N=2}=
2^{2-\mu}\ b^{k;N=2}_{aa'}\ V_{O6_k}^i\ 
\int\limits_0^\infty dl}
in the $NS$ sector from the zero momentum state $(r,s)=(0,0)$ will be discussed in a moment.

\subsec{Tadpole cancellation in one--loop gauge corrections}

One important question in any type $I$ one--loop calculation
is the possible existence of \UV divergences of the integrals \schwinger\ in the open
string channel $t\ra 0$. 
For the one--loop vacuum amplitude, given by the sum of \Ages, \Mges\ and the Klein bottle 
contribution, the cancellation of \UV divergences is guaranteed by the imposed tadpole 
cancellation.
The question of absence of \UV divergences in the full expression \oneloop\ has to be 
addressed again.
The background $B$, introduced in the open string partition function,
may give rise to NS--tadpoles of the graviton, dilaton and two--index 
antisymmetric tensor through couplings in the Born--Infeld action.
Furthermore, the R--piece of the \UV divergence may give rise to 
tadpoles of the Wess--Zumino type world brane couplings.
These  two types of divergences must vanish seperately in a consistent theory.

Eventually we would like all coefficients in front of the divergent integrals encountered before
to add up to zero. Therefore, we investigate the coefficients of the integrals and define
$\delta_X=\Tr(Q_a)^2\ \kappa_X \int\limits_0^\infty dl$, with $X$ being one of the subscripts.
Let us first collect the coefficients $\kappa$ of all potential divergent terms $\delta$ 
encountered in subsection 3.2 and refering to N=1 supersymmetric sectors. 
We have \followint
\eqn\closerlook{\eqalign{
\kappa_{ab}&=2i\ N_b\ I_{ab}\lf[\coth(\pi v^1_{ab})+\coth(\pi v^2_{ab})+\coth(\pi v^3_{ab})\ri]\cr
&=-2\ N_b\hskip-4mm\sum_{(i,j,l)=\atop (1,2,3),(2,1,3),(3,1,2)}\fc{1}{\sin\al^i}
\lf[n_a^in_b^i\fc{R_1^i}{R_2^i}+m_a^im_b^i\fc{R_2^i}{R_1^i}+(n_a^im_b^i+n_b^im_a^i)\cos\al^i\ri]
\ I_{ab}^j\ I_{ab}^l\cr
&=-N_b\hskip-4mm\sum_{(i,j,l)=\atop (1,2,3),(2,1,3),(3,1,2)}\hskip-4mm
\fc{1}{U_2^i}\lf[(n_a^i+m_a^iU^i)(n_b^i+m_b^i\ov U^i)+
(n_a^i+m_a^i\ov U^i)(n_b^i+m_b^iU^i)\ri]\ I_{ab}^j\ I_{ab}^l\ ,}}
and a similar expression from \followintmirror:
\eqn\closerlooka{\eqalign{
\kappa_{aa'}&=4i\ N_a\ I_{aa'}\lf[\coth(\pi v^1_{aa'})+\coth(\pi v^2_{aa'})
+\coth(\pi v^3_{aa'})\ri]\cr
&=-4\ N_a\hskip-4mm\sum_{(i,j,l)=\atop (1,2,3),(2,1,3),(3,1,2)}\hskip-4mm\fc{1}{\sin\al^i}
\lf[n_a^in_{a'}^i\fc{R_1^i}{R_2^i}+m_a^im_{a'}^i\fc{R_2^i}{R_1^i}+(n_a^im_\ap^i+n_\ap^im_a^i)
\cos\al^i\ri]I_{aa'}^j\ I_{aa'}^l\cr
&=-2N_a\hskip-8mm\sum_{(i,j,l)=\atop (1,2,3),(2,1,3),(3,1,2)}\hskip-6mm
\fc{1}{U_2^i}\lf[(n_a^i+m_a^iU^i)(n_a^i+m_a^i\ov U^i)+
(n_\ap^i+m_\ap^i\ov U^i)(n_\ap^i+m_\ap^iU^i)\ri]I_{a\ap}^jI_{a\ap}^l}}
Finally the relevant coefficient of \followintm\ may be rewritten:
\eqn\closerlookm{\eqalign{
\kappa^{k}_{a,\Om\Rc\th^ka}&=
-8i\ N_a\ \rho_{\Om\Rc\th^k}\ I_{a;O6_k}\ \lf[\coth(\pi v^{1;k}_a)+\coth(\pi v^{2;k}_a)+
\coth(\pi v^{3;k}_a)\ri]\cr
&=8\ N_a\ \rho_{\Om\Rc\th^k}\hskip-2mm\sum_{(i,j,l)=\atop (1,2,3),(2,1,3),(3,1,2)}\hskip-2mm
\fc{1}{\sin\al^i}\lf[n_{O6_k}^in_a^i\fc{R_1^i}{R_2^i}+m_{O6_k}^im_a^i\fc{R_2^i}{R_1^i}\ri.\cr
&\ \ \ \lf.+\ (n_a^im_{O6_k}^i+n_{O6_k}^im_a^i)\cos\al^i\ri]\ I^j_{a;O6_k}\ I^l_{a;O6_k}\cr
&=4\ N_a\ \rho_{\Om\Rc\th^k}\ \sum_{(i,j,l)=\atop (1,2,3),(2,1,3),(3,1,2)}
\fc{1}{U_2^i}\lf[(n_a^i+m_a^iU^i)(n_{O6_k}^i+m_{O6_k}^i\ov U^i)\ri.\cr
&\ \ \ \lf.+(n_a^i+m_a^i\ov U^i)(n_{O6_k}^i+m_{O6_k}^iU^i)\ri]\ I^j_{a;O6_k}\ I^l_{a;O6_k}\ .}}
The first observation we should make here is, that the tadpole  contributions 
\closerlook, \closerlooka, and \closerlookm\ boil down to the N=2 expressions 
\followintii,\ \divaa\ and \divm\ after respecting the intersection properties in that case.
That is why we have dropped the subscripts N=1 at the $\delta$'s in the above formulae. We 
shall use the latter in the following both for N=1 and N=2 sectors.
Moreover, from the expression in appendix \appC\ we realize, that those parts of 
the M\"obius $M_{aa'}$ and the annulus diagram $A_{ab}$, which give 
rise to one--loop gauge corrections, may have only  $NS$--tadpoles, but no $R$--tadpoles.
This statement takes over to the annulus $A_{aa'}$ diagram if only charges from one 
open string contribute (planar annulus), what precisley is the case in the other 
two diagrams mentioned before.
We have seen from \expandi, that only charges from one string end contribute in the case 
of anomlay free gauge groups, \ie in the case \eqq \anomalyfree\ holds.
Hence, the absence of $R$--tadpoles in one--loop gauge corrections is directly related
to the absence of gauge anomalies. On the other hand, in the case of
anomalous gauge groups in the non--planar annulus contribution of $\Delta_{aa'}$ an 
\UV divergence, proportional to $\Tr(Q)$, appears from tadpoles of an anti--symmetric tensor 
field.  
This field couples to the background gauge field through a Green--Schwarz interaction.
The situation is similar to what happens in the one--loop $F^6$ couplings of 
$SO(32)$ type $I$ string in $D=10$:
There the \UV divergence of a planar annulus diagram is cancelled against the divergence
from the M\"obius diagram. No potential gauge anomaly arises from these diagrams.
On the other hand, the anomaly of the non--planar annulus diagram is cancelled through
an exchange of an anti--symmetric tensor field in the closed string channel.
This effect is seen as massles closed string pole in the non--planar annulus calculation.

All the coefficients of potential $NS$--contributions \closerlook, \closerlooka\ and
\closerlookm\ have to be summed up according to \oneloop\ and give a zero result:
\eqn\oneloopzn{
\sum_{b=1\atop b\neq a}^K\sum_{a \in [a]\atop b\in [b]}\kappa_{ab}+
\sum_{a,a'\in [a]\atop a\neq a'} \kappa_{aa'}+\sum_{a,a'\in [a]\atop a\neq a'}\sum_{k=0}^{N-1}\ 
\kappa^k_{a,\Om\Rc\th^ka}=0\ .}
This expression has to vanish in order that no $NS$--tadpoles appear in our one--loop
gauge threshold calulation.
In the following let us prove the validity of this equation for the $\IZ_2\times \IZ_2$ 
orientifold.

\br
{$\IZ_2\times \IZ_2$ orientifold:}

As a concrete example, let us check the validity of \oneloopzn\ for the $\IZ_2\times \IZ_2$ 
orientifold.
Type $IIA$ compactified on a $\IZ_2\times \IZ_2$ orbifold, supplemented 
with the $\Om\Rc$ orientifold projection \reflection, represents a rather simple background for
intersecting branes to fulfill  the requirements for vacuum tadpole cancellation 
and N=1 chiral supersymmetry in $D=4$. These conditions are given by
{\it Eqs.} \twistedtadpoles\ and (3.49). For details see Ref. \doubref\cvetic\cveticaa.
The orbifold group is
represented by the twists $\th=\h(1,-1,0)$ and $\om=\h(0,1,-1)$.
We have the four $O6$--planes $\Om\Rc, \Om\Rc\th, \Om\Rc\om$, and $\Om\Rc\th\om$. 
Their $RR$--charges are cancelled by introducing stacks $a$ of $D6$ branes with
wrapping numbers $(n_a^j,m_a^j)$ w.r.t. to the three internal tori $T_2^j$.
The orbifold group generators $\th$ and $\om$ map each brane onto itself.
In that case to each stack $a$ only one mirror $a'=\Rc a$ is required. This means, that 
the conjugacy class of $a$ consists of only two elements, namely: $[a]=\{a,\Rc a\}$. 
Due to the simple structure of conjugacy classes the general formula \oneloop\ for the one--loop
gauge corrections to the gauge group $G_a$ boils down to:
\eqn\oneloopztwo{
\Delta_{G_a}=\sum_{b=1\atop b\neq a}^K\lf(\Delta_{ab}+\Delta_{ab'}+\Delta_{a'b}+\Delta_{a'b'}\ri)
+\Delta_{aa'}
+2(\Delta^{1}_{aa'}+\Delta^{\th}_{aa'}+\Delta^{\om}_{aa'}+\Delta^{\th\om}_{aa'})\ .}
Note, that there a four possible twist insertions $1,\th,\om,\th\om$ 
into the M\"obius diagram describing a string starting on a brane from stack $a$ and 
ending on its orientifold mirror $a'$. 

We shall investigate the cancellation of tadpoles in the gauge threshold result \oneloopztwo\
for any intersecting brane setup which fulfills the requirements of vacuum tadpole
cancellation. We first specialize to untilted tori $T_2^j$, \ie the latter are chosen to be 
rectangular ($\al^j=\fc{\pi}{2}$), though the generalization
to tilted two--tori is straightforward and will be discussed at the end of this subsection.
In that case, the orientifold mirror $a'$ of a brane $a$ with wrapping numbers $(n_a^j,m_a^j)$
takes the wrapping numbers $(n_a^j,-m_a^j)$.
According to \closerlook\ an N=1 annulus diagram, which describes a string starting on stack $a$, 
referring to the gauge group $G_a$ under consideration and which ends on stack $b$ 
from a different gauge group, contributes the coefficient
\eqn\di{\eqalign{
\sum^K_{b=1 \atop b\neq a}\sum_{a\in \{a,\Om\Rc a\}\atop b\in \{b,\Om\Rc b\}}
\kappa_{ab}&=\sum^K_{b=1 \atop b\neq a}
\kappa_{ab}+\kappa_{ab'}+\kappa_{a'b}+\kappa_{a'b'}\cr
&=-2\sum_{b=1\atop b\neq a}^K\sum_{b\in \{b,\Om\Rc b\}\atop a\in \{a,\Om\Rc a\}}
N_b\sum_{(i,j,k)=\atop (1,2,3),(2,1,3),(3,1,2)}\hskip-3mm
\lf(n_a^in_b^i\fc{R_1^i}{R_2^i}+m_a^im_b^i\fc{R_2^i}{R_1^i}\ri)\ I_{ab}^jI_{ab}^k\cr
&=8\lf[
(-n_a^1n_a^2n_a^3\fc{R_1^1}{R_2^1}+m_a^1m_a^2n_a^3\fc{R_2^2}{R_1^2}+
m_a^1n_a^2m_a^3\fc{R_2^3}{R_1^3})\sum_{b=1\atop b\neq a}^K N_b\ n_b^1m_b^2m_b^3\ri.\cr
&+(m_a^1m_a^2n_a^3\fc{R_2^1}{R_1^1}-n_a^1n_a^2n_a^3\fc{R_1^2}{R_2^2}+n_a^1m_a^2m_a^3
\fc{R_2^3}{R_1^3})\sum_{b=1\atop b\neq a}^K N_b\ m_b^1n_b^2m_b^3\cr
&+(m_a^1n_a^2m_a^3\fc{R_2^1}{R_1^1}+n_a^1m_a^2m_a^3\fc{R_2^2}{R_1^2}-n_a^1n_a^2n_a^3
\fc{R_1^3}{R_2^3})\sum_{b=1\atop b\neq a}^K N_b\ m_b^1m_b^2n_b^3  \cr
&\lf.-(n_a^1m_a^2m_a^3\fc{R_1^1}{R_2^1}+m_a^1n_a^2m_a^3\fc{R_1^2}{R_2^2}+m_a^1m_a^2n_a^3
\fc{R_1^3}{R_2^3})\sum_{b=1\atop b\neq a}^K N_b\ n_b^1n_b^2n_b^3\ri]}}
in front of the potential \UV divergence. 
Furthermore, an annulus diagram within an conjugacy class describes an open string exchange 
between stack $a$ and its mirror $a'$ and leads to tadpole coefficient:
\eqn\dii{\eqalign{
\kappa_{aa'}&=
-4N_a\sum_{(i,j,k)=\atop (1,2,3),(2,1,3),(3,1,2)}\lf(n_a^in_{a'}^i\fc{R_1^i}{R_2^i}
+m_a^im_{a'}^i\fc{R_2^i}{R_1^i}\ri)\ I_{aa'}^j\ I_{aa'}^k\cr
&=-16N_a\lf[m_a^2m_a^3(n_a^1)^2n_a^2n_a^3\fc{R_1^1}{R_2^1}+m_a^1m_a^3n_a^1(n_a^2)^2n_a^3
\fc{R_1^2}{R_2^2}+m_a^1m_a^2n_a^1n_a^2(n_a^3)^2\fc{R_1^3}{R_2^3}\ri.\cr
&-\lf.(m_a^1)^2m_a^2m_a^3n_a^2n_a^3\fc{R_2^1}{R_1^1}-m_a^1(m_a^2)^2m_a^3n_a^1n_a^3
\fc{R_2^2}{R_1^2}-m_a^1m_a^2(m_a^3)^2n_a^1n_a^2\fc{R_2^3}{R_1^3}\ri]\ .\cr}}
Finally, with the intersection numbers \interztwo\ and the general expression \closerlookm\
for the M\"obius divergence we obtain:
\eqn\Closerlookm{\eqalign{
\kappa_{aa'}^1&=
-2^6\ i\ \rho_{\Om\Rc}\ m_a^1m_a^2m_a^3\ \lf[\coth(\pi v^1_a)+\coth(\pi v^2_a)+
\coth(\pi v^3_a)\ri]\cr
&= 2^6\ \rho_{\Om\Rc}\ \lf(n_a^1m_a^2m_a^3\fc{R_1^1}{R_2^1}+ m_a^1n_a^2m_a^3\fc{R_1^2}{R_2^2} 
+ m_a^1m_a^2n_a^3\fc{R_1^3}{R_2^3}\ri)\ ,\cr
\kappa_{aa'}^\th&=
-2^6\ i\ \rho_{\Om\Rc\th}\ n_a^1n_a^2m_a^3\lf[\tanh(\pi v^1_a)+\tanh(\pi v^2_a)+
\coth(\pi v^3_a)\ri]\cr
&= 2^6\ \rho_{\Om\Rc\th}\ \lf(-m_a^1n_a^2m_a^3\fc{R_2^1}{R_1^1}- 
n_a^1m_a^2m_a^3\fc{R_2^2}{R_1^2} + n_a^1n_a^2n_a^3\fc{R_1^3}{R_2^3}\ri)\ ,\cr
\kappa_{aa'}^\om&=
-2^6\ i\ \rho_{\Om\Rc\om}\ m_a^1n_a^2n_a^3\ 
\lf[+\coth(\pi v^1_a)+\tanh(\pi v^2_a)+\tanh(\pi v^3_a)\ri]\cr
&=2^6\ \rho_{\Om\Rc\om}\ \lf(n_a^1n_a^2n_a^3\fc{R_1^1}{R_2^1}- m_a^1m_a^2n_a^3\fc{R_2^2}{R_1^2} 
- m_a^1n_a^2m_a^3\fc{R_2^3}{R_1^3}\ri)\ ,\cr
\kappa_{aa'}^{\th\om}&=
-2^6\ i\ \rho_{\Om\Rc\th\om}\ n_a^1m_a^2n_a^3\ \lf[\tanh(\pi v^1_a)+\coth(\pi v^2_a)+
\tanh(\pi v^3_a)\ri]\cr
&=2^6\ \rho_{\Om\Rc\th\om}\ \lf(-m_a^1m_a^2n_a^3\fc{R_2^1}{R_1^1}+
n_a^1n_a^2n_a^3\fc{R_1^2}{R_2^2} - n_a^1m_a^2m_a^3\fc{R_2^3}{R_1^3}\ri)\ .}}
The phases resulting from the traces \tracess\ may be taken from \bonni:
$\rho_{\Om\Rc\th}=\rho_{\Om\Rc\om}=\rho_{\Om\Rc\th\om}=-1$, and $\rho_{\Om\Rc}=1$.
With this information we add these tadpole contributions according to \oneloopztwo\  
and find a vanishing result:
\eqn\tadpolesumup{
\delta_{G_a}=\kappa_{ab}+\kappa_{ab'}+\kappa_{a'b}+\kappa_{a'b'}+\kappa_{aa'}
+2(\kappa^{1}_{aa'}+\ \kappa^{\th}_{aa'}+\kappa^{\om}_{aa'}+\kappa^{\th\om}_{aa'})=0\ .}
To prove this identity, one only needs to apply the vacuum RR tadpole constraints \cvetic:
\eqn\combine{\eqalign{
\sum_{a=1}^K N_a\  n_a^1 n_a^2 n_a^3=16\ \ \ \ ,\ \ \ \ &
\sum_{a=1}^K N_a\ n_a^1 m_a^2 m_a^3=-16\ \ \ ,\cr
\sum_{a=1}^K N_a\ m_a^1 n_a^2 m_a^3=-16\ \ \ \ ,\ \ \ \ &
\sum_{a=1}^K N_a\ m_a^1 m_a^2 n_a^3=-16\ .}}
In addition the twisted tadpole conditions \twistedtadpoles\ have to hold, which enabled
us to disregard $\th,\om,\th\om$--insertion in the annulus diagrams.
The generalization to tilted tori is straightforward. Only the following modifications have 
to be performed in the \eqqs \di--\Closerlookm: changing
$m_a^j\ra\tilde m_a^j=m_a^j+\h n_a^j$ in the case of a tilted torus $T_2^j$, and 
replacing in \eqq \Closerlookm\
the intersecting numbers for the untilted case with the relevant numbers \interztwo\  for
the tilted case. Finally, to proof \tadpolesumup\ for the tilted case, the vacuum tadpole condition
for that case \cveticaa\ has to be borrowed.
In fact, we went through all these steps to prove \tadpolesumup\ also for the tilted case.
Hence in the $\IZ_2\times \IZ_2$ orbifold/orientifold with intersecting branes, 
the cancellation of vacuum tadpole contributions \combine\ implies
the cancellation of tadpoles arising in an one--loop gauge threshold calculation.

\subsec{General structure of one--loop gauge threshold corrections}

We have already explained in subsection 3.1, that due to the orbifold/orientifold action a 
stack $a$ of $N_a$ branes has several mirrors, all together summarized in the conjugacy 
class $[a]$.
In orbifold/orientifold backgrounds with intersecting branes the spectrum and 
group representations of the open strings 
is organized according to which branes the open string ends couple.

From this analysis three different N=1 open string sectors are possible.
The open string sectors $aa,a'a',\ldots$, with $a,a',\ldots\in [a]$, which describe open strings
stretched between branes from stack $a,a',\ldots$ respectively, represent 
the vector multiplets describing the gauge group $G_a$. The latter
is subject to the orbifold and orientifold action. In addition there are adjoint chiral matter 
fields coming from this sector.
Since these strings are always stretched between two parallel branes, \ie preserve N=4
space--time supersymmetry, they are not relevant for gauge coupling renormalization.
Open strings stretched between one brane $a$ of the gauge group $G_a$ and an other brane $b$
of a different gauge group $G_b$ give rise to $I_{ab}$ chiral fermions in the bifundamental 
representation of the gauge groups $G_a$ and $G_b$. Here $I_{ab}$ is the intersection number 
\intersection\ of brane $a$ and $b$. 
More generally, after taking into account the various images contained in $[a]$ 
and $[b]$ we obtain from the open string sectors $ab+ba, ab'+b'a,a'b+ba',\ldots$
chiral fermions in the bifundamental of the groups $G_a$ and $G_b$ with multiplicities
$I_{ab},I_{ab'},I_{a'b},\ldots$, respectively.
We have seen, that those open string sectors contribute\foot{We do not display the potential 
divergences $\delta$ encountered in the previous subsections, as we have
shown in subsection 3.4, that they are cancelled anyway after adding up all 
those potential tadpole contributions (\cf \eqq \oneloopzn).} to the one--loop gauge correction
$\Delta_{G_a}$ the amount (\cf \altertiinew)
\eqn\summer{
\Delta^{\rm N=1}_{ab}=
-b_{ab}^{N=1}\  \ln \fc{\Gamma(1-\fc{1}{\pi}\phi_{ba}^1)\ 
\Gamma(1-\fc{1}{\pi}\phi_{ba}^2)\ \Gamma(1-\fc{1}{\pi}\phi_{ba}^3)}
{\Gamma(1+\fc{1}{\pi}\phi_{ba}^1)\ \Gamma(1+\fc{1}{\pi}\phi_{ba}^2)\ 
\Gamma(1+\fc{1}{\pi}\phi_{ba}^3)}  \ ,}
with $\phi_{ba}^j$ being the angles between the branes $a$ and $b$. The latter 
respect\foot{Note, that according to the comments made before \eqq \nontrivial\ 
the angles $\phi_{ba}^j$ are assumed to be shifted back into the range $0<|\phi_{ba}^j|<\pi$.}
the N=1 constraint $\phi_{ba}^1+\phi_{ba}^2+\phi_{ba}^3=0$. The coefficient (\cf \BETAA)
\eqn\bifund{
b_{ab}^{N=1}=N_b\ I_{ab}\ \Tr(Q_a^2)}
represents up to a sign the one--loop $\beta$--function 
coefficient\foot{The field--theoretical one--loop $\beta$--function coefficient is obtained
from the $IR$--limit $t\ra \infty$ of the integrands \schwinger\ after converting them 
into the open string channel. For $\Delta^{\rm N=1}_{ab}$ this leads to: 
\eqn\tischw{\Delta^{\rm N=1}_{ab}=\pi^{-1} i\ b_{ab}^{N=1}\ \int_0^\infty  \fc{dt}{t}\ \lf[
\fc{\th_1'(\fc{v^1_{ab}t}{2},\fc{it}{2})}{\th_1(\fc{v^1_{ab}t}{2},\fc{it}{2})}+
\fc{\th_1'(\fc{v^2_{ab}t}{2},\fc{it}{2})}{\th_1(\fc{v^2_{ab}t}{2},\fc{it}{2})}+
\fc{\th_1'(\fc{v^3_{ab}t}{2},\fc{it}{2})}{\th_1(\fc{v^3_{ab}t}{2},\fc{it}{2})}\ri]\ .}
In the limit $t\ra \infty$ the latter expression reproduces the one--loop running 
of the effective field theory (\cf \weinberg). 
With $\lim\limits_{t\ra\infty}\fc{\th_1'(\fc{vt}{2},\fc{it}{2})}{\th_1(\fc{vt}{2},\fc{vt}{2})}=
-(2k-1)\pi i\ , \ (k-1)<-iv<k\ ,$ 
we verify $b_{ab}^{N=1}$ as being related to the correct $\beta$--function coefficient 
$\beta_{ab}^{N=1}$ through:
\eqn\betarelation{
\beta_{ab}^{N=1}=\cases{-b_{ab}^{N=1}\ ,&$0<\phi^1_{ba}+\phi^2_{ba}<\pi\ ,$\cr
                         b_{ab}^{N=1}\ ,&$\pi<\phi^1_{ba}+\phi^2_{ba}<2\pi\ ,$}}
for the choice $0<\phi^1_{ba},\phi^2_{ba}<\pi$. Note, that the sign flip
in \betarelation\ is manifest in the logarithmic expression of \summer.} $\beta_{ab}^{N=1}$. 
It precisely accounts for $I_{ab}$ bifundamental representations 
$(N_a,\ov N_b),(N_a,N_b),\ldots$ of the gauge groups $U(N_a)\times U(N_b)$. 
The latter is the gauge group arising
from the stacks $a$ and $b$ before the orbifold and orientifold twists
(\cf also footnote 4 for the case, if one stack is parallel to an orientifold plane).
Note, that a negative intersection number $I_{ab}$ gives rise to $-I_{ab}$ fermions of opposite 
chirality. Finally, open strings starting and ending on two different branes from one 
conjugacy class $[a]$,
give rise to the sectors $aa'+a'a$, with $a,a'\in [a]$. We obtain $I_{aa'}$ 
symmetric and antisymmetric representations of $G_a$ from an open string stretched
between the branes $a$ and $a'$. 
Additional symmetric and antisymmetric representations of $G_a$ with multiplicity $I_{O6_k;a}$ 
arise, if $a'=\Om\Rc\th^k a$.
These sectors have annulus and M\"obius diagrams contributing to the one--loop
gauge correction $\Delta_{G_a}$, given by \altertiiaa\ and \altertiimm, respectively.
With the relevant choice of angles $\phi^j_{a'a}$ and $\phi^{j;k}_{a}$ 
these corrections take the same form as \summer\ up to a modification of 
their coefficients\foot{For the M\"obius diagram the transformation 
of \tiim\ back into the open string sector yields:
\eqn\timschw{\Delta^{k;\rm N=1}_{a,\Om\Rc\th^ka}=
\pi^{-1} i\ b_{a,\Om\Rc\th^ka}^{k;N=1}\ \int_0^\infty  \fc{dt}{t}\ \lf[
\fc{\th_1'(v^{k;1}_{a}t,\fc{it}{2}+\h)}{\th_1(v^{k;1}_{a}t,\fc{it}{2}+\h)}+
\fc{\th_1'(v^{k;2}_{a}t,\fc{it}{2}+\h)}{\th_1(v^{k;2}_{a}t,\fc{it}{2}+\h)}+
\fc{\th_1'(v^{k;3}_{a}t,\fc{it}{2}+\h)}{\th_1(v^{k;3}_{a}t,\fc{it}{2}+\h)}\ri]\ .}
Hence the field--theoretical $\beta$--function coefficient is equal to 
$b^{N=1}_{a;\Om\Rc\th^k a}$ up to a sign
\eqn\betarelationm{
\beta_{a;\Om\Rc\th^k a}^{N=1}=
\cases{-b_{a;\Om\Rc\th^k a}^{N=1}\ ,&$0<\phi^{k;1}_{a}+\phi^{k;2}_{a}<\fc{\pi}{2}\ ,$\cr
        b_{a;\Om\Rc\th^k a}^{N=1}\ ,&$\fc{\pi}{2}<\phi^{k;1}_{a}+\phi^{k;2}_{a}<\pi\ ,$}}
for the choice $0<\phi^{k;1}_{a},\phi^{k;2}_{a}<\pi$.}
(\cf \eqqs \betaaa\ and \BETAm):
\eqn\antisymm{\eqalign{
b_{aa'}^{N=1}&=2\ N_a\ I_{aa'}\ \Tr(Q_a^2)\ ,\cr
b_{a;\Om\Rc\th^k a}^{k;N=1}&=-2\ N_a\ \rho_{\Om\Rc\th^k}\ I_{a;O6_k}\ \Tr(Q_a^2)}}
accounting for the number of possible symmetric and antisymmetric representations of
the respective sector.

Two branes, which are parallel w.r.t. to one torus $T_2^i$, but have 
non--trivial intersections w.r.t. to the remaining two tori $T_2^j,T_2^l$, 
preserve N=2 supersymmetry. Hence open string sectors associated to such branes give 
rise to N=2 vectormultiplets and hypermultiplets.
From the sectors $ab+ba, ab'+b'a,a'b+ba',\ldots$ we obtain respectively
$I^j_{ab}I^l_{ab},I^j_{ab'}I^l_{ab'},I^j_{a'b}I^l_{a'b},\ldots$ copies of
hypermultiplets in the bifundamental of the groups $G_a$ and $G_b$.
To the full correction $\Delta_{G_a}$ these sectors constitute the one--loop gauge correction:
\eqn\summerii{
\Delta^{\rm N=2}_{ab}=-b_{ab}^{N=2}\ \lf[\ln T^i_2V_a^i|\eta(T^i)|^4-\kappa\ri]\ .}
Here, $V_a^i$ represents the wrapped brane volume \newmoduli
\eqn\wrappedv{
V_a^i=\fc{1}{U_2^i}|n_a^i+U^im_a^i|^2\ ,}
w.r.t. to the torus $T_2^i$ and the K\"ahler modulus $T^i$ is defined in \torusmoduli.
The one--loop $\be$--function coefficient is given by (\cf \eqq \betaii):
\eqn\susreff{
b_{ab}^{N=2}=-2N_b\ I^j_{ab}\ I^l_{ab}\ \Tr(Q_a^2)\ ,}
accounting for $I^j_{ab}\ I^l_{ab}$ bifundamental representations $(N_a,\ov N_b)$ of
the gauge group $U(N_a)\times U(N_b)$. This coefficient agrees with the field--theoretical
N=2 one--loop beta--function coefficient.
Furthermore, the one--loop corrections from the open string sector $aa'+a'a$ 
take a similar form, with the $\beta$--function coefficients:
\eqn\antisymmm{\eqalign{
b_{aa'}^{N=2}&=-4\ N_a\ I^j_{aa'}\ I^l_{aa'}\ \Tr(Q_a^2)\ ,\cr
b_{a;\Om\Rc\th^k a}^{k;N=2}&=8\ N_a\ \rho_{\Om\Rc\th^k}\ I^j_{a;O6_k}\ I^l_{a;O6_k}\Tr(Q_a^2)\ .}}

According to \oneloop\ the one--loop gauge threshold correction $\Delta_{G_a}$ to the gauge 
group $G_a$ is organized as a sum over all possible individual gauge corrections 
$\Delta_{ab},\Delta_{aa'}$ and $\Delta^k_{a;\Om\Rc\th^ka}$ originating from the various 
open string sectors encountered before.
Apart from the topological numbers \bifund, \antisymm, \susreff\ and \antisymmm\  
these corrections take a rather universal form, given by \summer\ and \summerii. 
Hence the latter represent the two basic building blocks for the one--loop gauge 
correction $\Delta_{G_a}$, valid for any orbifold/orientifold backgrounds.
It is quite reassuring, that the numbers \BETAA, \BETAm\ and 
\betaaa\ arising from the string calculation precisely match (up to a sign) 
the field--theoretical expressions. 
The same applies for the N=2 (string--theoretical) $\be$--function
coefficients \betaii, \betamiia\ and \betamii.
To conclude, the gauge threshold correction $\Delta_{G_a}$ is fully determined 
by the two basic functions \summer\ and \summerii, the model dependent angles and moduli
entering those functions and the field--theoretical $\beta$--function coefficients.

This is in close analogy to the results on the heterotic side \DKL\ or for
type $I$ orientifolds with non--intersecting  $D9$ and $D5$  branes \doubref\BF\ABD.
On the other hand, from the close analogy  between one--loop gauge threshold corrections
from the N=2 sectors of intersecting branes, given by \summerii, and 
one--loop gauge threshold corrections
from the N=2 sectors of type $I$ orientifold compactifications,
one would have been tempted to speculate, that one--loop gauge
corrections from N=1 sectors are also moduli independent constants,
just as they are in type $I$ orbifold/orientifold compactifications with parallel branes 
\doubref\BF\ABD.
However, through \angle\ (or more precisely \Relation) the angles $\phi^j_{ab}$ imply a 
non--trivial radius $R_1^j,R_2^j$ dependence of \summer, which we shall uncover in the 
next subsection.
This leads to interesting moduli dependence of one--loop gauge corrections
for N=1 supersymmetric brane world models.

\subsec{Moduli dependence of N=1 gauge threshold corrections}

In ordinary orientifold compactifications, where the angles
$\phi^j=i\pi v^j$  take discrete values, the $\th$--functions 
in \ti\ do not give rise to any moduli dependence, and the N=1 threshold correction
$\Delta^{\rm N=1}_{ab}$ is just a constant, in agreement with the results of \doubref\BF\ABD.
This is obvious from our general result \altertiinew, which holds for arbitrary 
orientifold/orbifold backgrounds with intersecting branes. 
The same applies for \altertiimm\ and \altertiiaa.
However, in the case of branes at angles $\phi^j$, the
angles are given through the relation \angle. Therefore, the latter 
implies a non--trivial dependence of $\Delta^{\rm N=1}_{ab}$ 
on the radii $R_1^j,\ R_2^j$ encoded in \Relation.
We want to determine this dependence in this subsection.

Rather than starting with \nontrivial\ as before, we shall start with the relation 
\eqn\Nontrivial{
\fc{\th_1'(iv,\tau)}{\th_1(iv,\tau)}=-i\fc{\p}{\p v}\ln\th_1(iv,\tau)=
-i\fc{\pi}{F}-2\pi i\lf(F-\fc{1}{F}\ri)\ 
\sum_{k=1}^\infty F^{2k}\ C_{2k}(\tau)\ ,}
which can be derived from \eqq \nontrivial\ and:
\eqn\FFF{
\pi v=\ath\ F\ \ ,\ \ \fc{d F}{d v}=\pi(1-F^2)\ .}
Here, the functions $C_{2k}(\tau)$ are given by \tseytlin
\eqn\functions{
C_{2k}(\tau)=\sum_{n=1}^\infty\lf(\fc{1+q^n}{1-q^n}\ri)^{2k}=\sum_{n=1}^k\ c(k,n)\ (1-E_{2n})\ ,}
with the coefficients:
\eqn\strangefunctions{\eqalign{
c(1,1)&=\fc{1}{6}\ ,\cr
c(2,1)&=\fc{2}{9}\ ,\ \ \ c(2,2)=-\fc{1}{90}\ ,\cr
c(3,1)&=\fc{23}{90}\ ,\ \ c(3,2)=-\fc{1}{45}\ ,\ c(3,3)=\fc{1}{945}\ ,\cr
c(4,1)&=\fc{88}{315}\ ,\ c(4,2)=-\fc{22}{675}\ ,\ c(4,3)=\fc{8}{2835}\ ,\ 
c(4,4)=-\fc{1}{9450}\ ,\ \ \ \ldots\ .}}
Inserting \Nontrivial\ into \ti\ we obtain:
\eqn\tiinew{
\Delta^{\rm N=1}_{ab}=\delta_{ab}^{\rm N=1}
+4i\ b_{ab}^{N=1}\ \int_0^\infty  dl\ \sum_{j=1}^3 
\lf(F_{ab}^j-\fc{1}{F_{ab}^j}\ri)
\sum_{k=1}^\infty (F_{ab}^j)^{2k}\ C_{2k}(\tau)\ ,}
with the background gauge fields
\eqn\Relationn{\eqalign{
F_{ab}^j&=\tanh(\pi v_{ab}^j)=-i\tan(\phi_b^j-\phi_a^j)\cr
&=-\ I_{ab}^j\ \fc{U^j-\ov U^j}{(n_a^j+m_a^jU^j)(n_b^j+m_b^j\ov U^j)+
(n_a^j+m_a^j\ov U^j)(n_b^j+m_b^jU^j)}\ ,}}
and the $UV$--divergent part $\delta_{ab}^{\rm N=1}$, given in \followint, and eventually cancelled
in the whole result \oneloop.
Using the integrals \form\ we obtain:
\eqn\tiinewi{\eqalign{
\Delta^{\rm N=1}_{ab}&=\delta_{ab}^{\rm N=1}-4\pi i \ b_{ab}^{N=1}\cr
&\times\sum_{j=1}^3
\lf(F_{ab}^j-\fc{1}{F_{ab}^j}\ri)\sum_{k=2}^\infty (F^j_{ab})^{2k}\ 
\sum_{n=2}^{k}\ \fc{(2\pi)^{-2n}(2n)!}{(1-2n)|B_{2n}|}\ c(k,n)\ \zeta(2n-1)\cr
&=\delta_{ab}^{\rm N=1}+\fc{2i}{3\pi^3}\ b_{ab}^{N=1}\ \zeta(3)\ 
\sum_{j=1}^3 (F^j_{ab})^3\cr
&+4\pi i\ b_{ab}^{N=1}\ \sum_{k=2}^\infty\ 
\sum_{j=1}^3 (F^j_{ab})^{2k+1}
\lf\{\sum_{n=2}^{k+1} \fc{(2\pi)^{-2n}(2n)!}{(1-2n)|B_{2n}|}\ \zeta(2n-1)\ 
[c(k+1,n)-c(k,n)]\ri\}.}}
Similar expressions can be derived for the other two corrections $\Delta^{\rm N=1}_{aa'}$
and $\Delta^{k;\rm N=1}_{a,\Om\Rc\th^k a}$.

The series in the expression \tiinewi\ shows some interesting form, which
may resemble its origin from $D=10$ type IIA string theory.
In fact, in the one--loop N=2 prepotential in $d=4$ (describing $F^2$ corrections \HM)
or the N=1 prepotential (describing $F^4$ corrections \LS) in $d=8$, 
there occurs a $\zeta(3)$ and $\zeta(5)$, respectively as a result
of dimensional reducing higher gravitational couplings from $d=10$.
Thus it is tempting to believe, that the coefficients $\zeta(2n-1)$ in the series 
\tiinewi\ also originate from higher gravitational couplings in $d=10$.
Furthermore, the latter prepotentials have a geometric meaning in the corresponding
dual string theory, which should also be the case here (\cf the remarks in the conclusion).

\subsec{Explicit results for the $\IZ_2\times \IZ_2$ orientifold}

In subsection 3.4 we have verified the absence of tadpoles in the one--loop gauge 
corrections \oneloop\ for the $\IZ_2\times \IZ_2$ orientifold with intersecting branes.
This proof assumed an arbitrary choice of intersecting brane environment
with the only condition on the latter, that it had to fulfill the vacuum tadpole conditions 
\combine.
In this subsection we take for the $\IZ_2\times \IZ_2$ orientifold a particular choice of 
intersecting branes, which obeys \combine. For this model we 
shall calculate the corrections \oneloop, which boil down to 
\oneloopztwo\ for the $\IZ_2\times \IZ_2$ orientifold.
More concretely, we shall study one of the concrete models presented in \cvetic.
This represents a special N=1 supersymmetric solution of the vacuum tadpole equations \combine\ 
with six stacks of $D6$ branes, whose wrapping numbers are displayed in the following table:
\vskip0.5cm
{\vbox{\ninepoint{
\def\ss#1{{\scriptstyle{#1}}}
$$
\vbox{\offinterlineskip\tabskip=0pt
\halign{\strut\vrule#
&~$#$~\hfil
&\vrule#
&~$#$~\hfil
&\vrule#&\vrule#
&~$#$~\hfil
&~$#$~\hfil
&~$#$~\hfil
&~$#$~\hfil
&~$#$~\hfil
&\vrule#
&~$#$~\hfil
&\vrule#
&~$#$~\hfil
&\vrule#  
\cr
\noalign{\hrule}
&
{\rm Stack}
&&
{\rm Gauge\ group}
&&&
\phi_a^1
&&
\phi_a^2
&&
\phi_a^3
&&
N_a
&&
{\rm SUSY}
&
\cr
\noalign{\hrule}
\noalign{\hrule}
&
1
&&
U(3)\times U(1)
&&&
\arctan(2U)
&&
-\arctan(2U)
&&
0
&&
6+2
&&
N=2
&
\cr
&
2
&&
U(1)
&&&
0
&&
0
&&
0
&&
2
&&
N=4
&
\cr
&
3
&&
USp(4)
&&&
0
&&
\arctan(2U)
&&
-\arctan(2U)
&&
4
&&
N=2
&
\cr
&
4
&&
USp(8)
&&&
0
&&
\fc{\pi}{2}
&&
-\fc{\pi}{2}
&&
8
&&
N=2
&
\cr
&
5
&&
U(1)
&&&
\arctan(4U)
&&
0
&&
-\arctan(4U)
&&
2
&&
N=2
&
\cr
&
6
&&
USp(8)
&&&
\fc{\pi}{2}
&&
0
&&
-\fc{\pi}{2}
&&
8
&&
N=2
&
\cr
\noalign{\hrule}}}$$
\vskip-10pt
\centerline{\noindent{\bf Table 1:}
{\sl $D6$ brane configuration: angles and supersymmetry}}
\centerline{\sl w.r.t. the orientifold plane $\Omega\Rc$.}
\vskip10pt}}}
The $D6$--branes are assumed to pass through fix points.
The requirement for N=1 supersymmetry leads to the choice of complex
structure moduli: $U:=U^1=2U^2=U^3$ in the three internal tori $T_2^i$.
Furthermore for an untilted torus $a$ we have $\mu=0$ in \zeromodesm.

As a concrete example we focus on the gauge group $USp(4)$
referring to the third stack, with $N_a=4$.
From the Table 1 we instantly see, that branes from that stack preserve at least N=2 supersymmetry 
with their orientifold mirrors $a'=\Om\Rc a$. Hence only the correction $\Delta_{aa'}^{N=2}$
has to be discussed for open string  (annulus) exchanges between branes within the conjugacy class 
$[a]=\{a,\Om\Rc a\}$. Using \aaprime, $I^j_{aa'}=-2m_a^jn_a^j$, \ie
$I_{aa'}^2=-4$ and $I_{aa'}^3=2$ we obtain:
\eqn\AI{
\Delta^{\rm N=2}_{aa'}=-32N_a\ \Tr(Q_{USp(4)}^2)\ 
\lf[\ln |\eta(iR^1_1R_2^1)|^4+2\ln(R_1^1)-\kappa\ri]\ .}
Let us now determine the contributions from the M\"obius sector.
Without twist insertion, all the diagrams $M^0_{a,\Om\Rc a}$ describe N=2 supersymmetric
open string exchanges from brane $a$ to its orientifold mirror $a'=\Om\Rc a$ 
and we directly get from \tmoeb
\eqn\TM{\eqalign{
\Delta^{0;\rm N=2}_{a,\Om\Rc a}&=16 N_a\  n_a^1m_a^2m_a^3 \ \Tr(Q_{USp(4)}^2)
\lf[2\ln R_1^1+\ln|\eta(iR_1^1R_2^1)|^4-\ln 2-\kappa\ri]\cr
&=-32N_a\ \Tr(Q_{USp(4)}^2)
\lf[2\ln R_1^1+\ln|\eta(iR_1^1R_2^1)|^4-\ln 2-\kappa\ri]\ .}}
Furthermore, a twist insertions $\om$ does not act in the first torus, which is responsible for
the zero mode contributions of \TM. Hence, in \TM\ we only have to change the intersection numbers 
referring to the second and third plane from $I_{aa'}^{0;j}=2m_a^j$ to $I_{aa'}^{\om;2}=2n_a^2$ 
and $I_{aa'}^{\om;3}=-2n_a^3$  to obtain:
\eqn\TMM{\eqalign{
\Delta^{\om;\rm N=2}_{a,\Om\Rc a}&=-16N_a\  n_a^1 n_a^2 n_a^3\ \Tr(Q^2_{USp(4)})
\lf[2\ln R_1^1+\ln|\eta(iR_1^1R_2^1)|^4-\ln 2-\kappa\ri]\cr
&=-16N_a\ \Tr(Q^2_{USp(4)})
\lf[2\ln R_1^1+\ln|\eta(iR_1^1R_2^1)|^4-\ln 2-\kappa\ri]\ .}}
On the other hand, the twist insertions $\th$ and $\th\om$ lead to N=1 M\"obius diagrams, 
which after \altertiimm\ assume the form:
\eqn\TMMM{\eqalign{
\Delta^{\th;\rm N=1}_{a,\Om\Rc a}&=
-8N_a\ n_a^1n_a^2m_a^3\ \Tr(Q^2_{USp(4)})\ln \fc{\Gamma(1-\fc{\phi_a^1}{\pi}+\h)
\Gamma(1-\fc{\phi_a^2}{\pi}-\h)\Gamma(1-\fc{\phi_a^3}{\pi})}{\Gamma(1+\fc{\phi_a^1}{\pi}-\h)
\Gamma(1+\fc{\phi_a^2}{\pi}+\h)\Gamma(1+\fc{\phi_a^3}{\pi})}\cr
&=8N_a\ \Tr(Q^2_{USp(4)})\ln \h\fc{
\Gamma[\h-\fc{1}{\pi}\at]\Gamma[1+\fc{1}{\pi}\at]}{
\Gamma[\fc{3}{2}+\fc{1}{\pi}\at]\Gamma[1-\fc{1}{\pi}\at]}\ ,\cr
\Delta^{\th\om;\rm N=1}_{a,\Om\Rc a}&=-8N_a\ n_a^1m_a^2n_a^3\ \Tr(Q^2_{USp(4)})\ln 
\fc{\Gamma(1-\fc{\phi^1_a}{\pi}+\h)\Gamma(1-\fc{\phi^2_a}{\pi})\Gamma(1-\fc{\phi_a^3}{\pi}-\h)}
{\Gamma(1+\fc{\phi_a^1}{\pi}-\h)\Gamma(1+\fc{\phi_a^2}{\pi})\Gamma(1+\fc{\phi_a^3}{\pi}+\h)}\cr
&=-16N_a\ \Tr(Q^2_{USp(4)})\ln \h\fc{
\Gamma[1-\fc{1}{\pi}\at]\Gamma[\h+\fc{1}{\pi}\at]}{
\Gamma[1+\fc{1}{\pi}\at]\Gamma[\fc{3}{2}-\fc{1}{\pi}\at]}\ .}}
Let us now come to the annulus contributions from open strings starting on branes $a$ or their 
mirrors from the stack constituing the gauge group $USp(4)$ and ending on branes $b$ or 
their mirrors from different stacks.
Summing up all N=2 sectors gives rise to
\eqn\AA{\eqalign{
\Delta_{ab}^{N=2}&=32N_a\ \Tr(Q^2_{USp(4)})\ [-6\ln R_1^1-2\ln R_1^2-2\ln R_2^2
-\ln\lf(\fc{R_1^2}{R_2^2}+4\fc{R_2^2}{R_1^2}\ri)\cr
&-3\ln|\eta(iR_1^1R_2^1)|^4-\ln|\eta(iR_1^2R_2^2)|^4+4\kappa]\ ,}}
whereas the N=1 sectors sum up to:
\def\xt{\fc{1}{\pi}\arctan(2U)}
\def\xf{\fc{1}{\pi}\arctan(4U)}
\def\g{\Gamma}
\eqn\AAA{\eqalign{
\Delta_{ab}^{N=1}&=
N_a\ \Tr(Q^2_{USp(4)})\ \lf\{256\ln\fc{\g[1-\xt]}{\g[1+\xt]}+32\ln\fc{\g[\h+\xt]}{\g[\h-\xt]}\ri.\cr
&+32\ln\fc{\g[\fc{3}{2}+\xt]}{\g[\fc{3}{2}-\xt]}
+64\ln\fc{\g[1+2\xt]}{\g[1-2\xt]}\cr
&+32\ln\fc{\g[1-\xf]}{\g[1+\xf]}
+16\ln\fc{\g[1+\xt-\xf]}{\g[1-\xt+\xf]}\cr
&\lf.+48\ln\fc{\g[1+\xt+\xf]}{\g[1-\xt-\xf]}\ \ri\}\ .}}
Finally summing up all seven contributions \TM--\AAA\ gives the one--loop gauge 
threshold correction 
$\Delta_{USp(4)}$ to the gauge group $USp(4)$:
\eqn\finalcv{
\Delta_{USp(4)}=2\Delta^{\th;\rm N=1}_{a,\Om\Rc a}+2\Delta^{\th\om;\rm N=1}_{a,\Om\Rc a}+
\Delta_{ab}^{N=1}+\Delta^{\rm N=2}_{aa'}+2\Delta^{0;\rm N=2}_{a,\Om\Rc a}+
2\Delta^{\om;\rm N=2}_{a,\Om\Rc a}+\Delta_{ab}^{N=2}\ .}

\goodbreak

\newsec{Conclusions}

We have calculated the one--loop corrections to gauge couplings in 
N=1 supersymmetric brane world models. These models are realized through stacks of $D6$ branes, 
which are placed into an orbifold/orientifold background of type $IIA$ string theory and 
wrapped on 3--cycles with non--vanishing intersections.
The one--loop gauge thresholds are organized as a sum \oneloop\ over the 
corrections coming from the individual open string sectors stretched between the various
intersecting $D6$--branes.
The correction associated to one of this sector takes a universal form given by 
the one--loop $\beta$--function coefficient multiplied by one of the two basic 
functions \summer\ and \summerii. 
The moduli dependence of this sector enters these functions through the respective angles
describing the open string sector (encoded in \Relation).
Hence the complete correction $\Delta_{G_a}$ is fully determined 
by the two functions \summer\ and \summerii, the model dependent angles and moduli
entering those functions and the field--theoretical $\beta$--function coefficients.

The supersymmetric orientifold models considered here map in the strong coupling limit 
to compactifications of $M$--theory on certain singular $G_2$ manifolds.
Thus our N=1 gauge threshold function \summer\ is possibly
related to the recently calculated Ray--Singer torsion of singular $G_2$ manifolds \FW.
Besides \summer\ may give a hint on the form of non--perturbative corrections
to the gauge couplings on the heterotic side. 
In view of Refs. \doubref\LSquarter\morales\ the correction \summer\ represents an 
other class of function, which describes one--loop couplings with only $1/4$ BPS states 
contributing and which maps 
to non--perturbative corrections on the dual heterotic side.
Hence the object entering the integrand \ti\ represents a (weighted) counting function of 
these states, which is  mapped on the dual heterotic side to a possible topological quantity, 
associated to the singular $G_2$ compactification manifold.

Since the angles are related to the radii of the type $IIA$ compactification through \Relationn\ 
the angles imply a non--trivial radius dependence, shown in \tiinewi. This has the consequence
that, in contrast to what is known from ordinary orbifold/orientifold theories, 
N=1 subsectors do give rise to moduli--dependent one--loop corrections, which may become 
huge for certain regions in 
the moduli space. This fact has an important impact on the unification scale
and other phenomenological properties in
intersecting brane world models \progress.

\bigskip
\centerline{\bf Acknowledgments }\nobreak
\bigskip

We wish to thank R. Blumenhagen, L. G\"orlich, G. Honecker, and T. Ott for 
numerous and valuable discussions.
This work is supported in part by the Deutsche Forschungsgemeinschaft (DFG), and 
the German--Israeli Foundation (GIF).
St.St. thanks the CERN Theory Division for hospitality during completion of this work.


\goodbreak
\appendix{\appA}{Modular functions and Dirichlet series}

In this appendix we want to investigate the following type of integrals
\eqn\investigate{
\int\limits_0^\infty  dy\ y^{s-1}\ [E_{2k}(iy)-1]}
as they appear in the calculation of one--loop gauge threshold corrections
for N=1 sectors (cf. \tiinew). Essentially we shall prove \eqq \form.
This leads us to the connection between a modular form 
with Fourier series
\eqn\FOURIER{
f(\tau)=c(0)+\sum_{n=1}^\infty c(n)\ e^{2\pi i n\tau}}
and the Dirichlet series 
\eqn\DIRICHLET{
D(s)=\sum_{n=1}^\infty\fc{c(n)}{n^s}\ ,}
which was established by Hecke, see \eg \APO.
Let us review the relevant steps.
If $f$ is an element of the set of all entire modular forms of weight $2k$,
its Fourier coefficients fulfill $c(n)\sim \Oc(n^{2k-1})$. 
Hence the series  \DIRICHLET\ is absolutely  convergent for $\re(s)>2k$.
For $k>1$ the isomorphism  between \FOURIER\ and \DIRICHLET\ is made transparent
through the integral: 
\eqn\transparent{
(2\pi n)^{-s}\ \Gamma(s)=\int\limits_0^\infty \ y^{s-1}\ e^{-2\pi n y}\ dy\ \ \ ,
\ \ \ \re(s)>0\ .}
With multiplying both sides by $c(n)$ and sum over $n$ we obtain:
\eqn\obtain{
(2\pi)^{-s}\ \Gamma(s)\ D(s)=\int\limits_0^\infty\ y^{s-1}\ [f(iy)-c(0)]\ dy\ .}
In the above step we exchanged the order of summation and integration, which
is only valid for $\re(s)>2k$.
However, after explicitly evaluating\foot{This makes use of the modular behaviour of 
$f$, which is guaranteed for $k>1$, what is our assumption. Note, that there does not 
exist a modular function of weight $2$.} 
the integral on the r.h.s.
\eqn\expliciteva{
(2\pi)^{-s}\ \Gamma(s)\ D(s)=\int\limits_1^\infty\ [y^s+(-1)^k\ y^{2k-s}]\ 
[f(iy)-c(0)]\ dy-c(0)\ \lf(\fc{1}{s}+\fc{(-1)^k}{2k-s}\ri)}
one may obtain an analytic continuation to the region $\re(s)<2k$  
with $D(s)$ fulfilling the following functional equation \APO:
\eqn\functional{
(2\pi)^{-s}\ \Gamma(s)\ D(s)=(-1)^{k}\ (2\pi)^{s-2k}\ \Gamma(2k-s)\ D(2k-s)\ .}
At $s=2k$ there is a pole with residue:
\eqn\residue{
\fc{(-1)^{k}\ (2\pi)^{2k}\ c(0)}{\Gamma(2k)}\ .}

The procedure outlined above may be applied for the Eisenstein series
\eqn\eisen{
E_{2k}(\tau)=1+\fc{(2\pi i)^{2k}}{(2k-1)!\ \zeta(2k)}\ \sum_{n=1}^\infty 
\sigma_{2k-1}(n)\ e^{2\pi i n\tau}\ \ \ ,\ \ \ k\geq 1\ ,}
with:
\eqn\SIGMAS{
\sigma_{2k-1}(n)=\sum\limits_{d=1\atop d|n}^\infty d^{2k-1}\ .}
The Dirichlet series associated to the function $f(\tau)=E_{2k}(\tau)$ 
may be obtained from \obtain\ for $\re(s)>2k$.
With  using \transparent\ we may evaluate the following integral for $s>0$
\eqn\following{\eqalign{
(2\pi)^{-s}\ \Gamma(s)\ D(s)&=\int\limits^\infty_0 y^{s-1}\ [E_{2k}(iy)-1]\ dy\cr
&=\fc{(2\pi i)^{2k}}{(2k-1)!\ \zeta(2k)}\ \Gamma(s)\ \sum_{n=1}^\infty \sigma_{2k-1}(n)
\ (2\pi n)^{-s}\cr
&=\fc{(2\pi i)^{2k}}{(2k-1)!\ \zeta(2k)}\ 
\Gamma(s)\ \sum_{m=1}^\infty\sum_{d=1}^\infty \ d^{2k-1}\ (2\pi dm)^{-s}\cr
&=\fc{(2\pi i)^{2k}}{(2k-1)!\ \zeta(2k)}\ (2\pi)^{-s}
\ \Gamma(s)\ \zeta(1+s-2k)\ \zeta(s)\ .}}
Thus, for $\re(s)>2k$  we obtain the Dirichlet series
\eqn\finalDirichlet{
D(s)=\fc{(2\pi i)^{2k}}{(2k-1)!\ \zeta(2k)}\ \zeta(1+s-2k)\ 
\zeta(s)\ \ ,\ \ \re(s)>2k\ ,}
associated to the modular function $E_{2k}$.
However, if in addition $k>1$, 
the relation \functional\ may be applied to give us the analytic continuation 
of \finalDirichlet\ to the region $\re(s)<2k$:
\eqn\continued{\eqalign{
D(s)&=(-1)^{k}\ (2\pi)^{2s-2k}\ \fc{\Gamma(2k-s)}{\Gamma(s)}\ D(2k-s)\cr
&=\fc{(2\pi i)^{2k}}{(2k-1)!\ \zeta(2k)}\ (-1)^{k}\ (2\pi)^{2s-2k}\ 
\fc{\Gamma(2k-s)}{\Gamma(s)}\ \zeta(1-s)\ \zeta(2k-s)\ \ ,\ \ \re(s)<2k\ .}}

The previous equations allow us to  extract an expression for the integral
\investigate:
\eqn\cern{\eqalign{
\int\limits_0^\infty  dy\ y^{s-1}\ [E_{2k}(iy)-1]&\cr
&\hskip-30mm=\cases{
\fc{(2\pi i)^{2k}}{(2k-1)!\ \zeta(2k)}\ (2\pi)^{-s}
\ \Gamma(s)\ \zeta(1+s-2k)\ \zeta(s)\ ,&$\re(s)>2k$\ ,  \cr
\fc{(2\pi i)^{2k}}{(2k-1)!\ \zeta(2k)}\ (-1)^{k}\ (2\pi)^{s-2k}\ 
\Gamma(2k-s)\ \zeta(1-s)\ \zeta(2k-s)\ ,&$\re(s)<2k$\ .\cr}}}
In particular, the case $s=1$ leads to ($k>1$):
\eqn\inparticular{\eqalign{
\int\limits_0^\infty  dy\ [E_{2k}(iy)-1]&=-\h\ 
\fc{(2\pi i)^{2k}}{(2k-1)!\ \zeta(2k)}\ (-1)^{k}\ (2\pi)^{1-2k}\ 
\Gamma(2k-1)\ \zeta(2k-1)\cr
&=\fc{\pi}{(1-2k)\zeta(2k)}\ \zeta(2k-1)\ .}}
This result may be also directly anticipated from regulating the integral \following.
Indeed replacing in \following\ the integer $s$ by $1+\eps$, with $\eps>0$ playing 
the role of a regulator, we obtain:
\eqn\rhs{\eqalign{
\int\limits_0^\infty  dy\ y^\eps\ [E_{2k}(iy)-1]&=
\fc{(2\pi i)^{2k}}{(2k-1)!\ \zeta(2k)}\ (2\pi)^{-1-\eps}\ \Gamma(1+\eps)\ 
\zeta(2+\eps-2k)\ \zeta(1+\eps)\cr
&\hskip-30mm=\cases{-\h\ 
\fc{(2\pi i)^{2k}}{(2k-1)!\ \zeta(2k)}\ (-1)^{k}\ (2\pi)^{1-2k}\ 
\Gamma(2k-1)\ \zeta(2k-1)+\Oc(\eps)\ ,&$k>1\ ,$\cr
\fc{6}{\pi \eps}+\Oc(\eps)\ ,&$k=1$\ .}}}
The case $k>1$, clearly agrees with \inparticular\ in the limit $\eps\ra 0$.
However, in addition \rhs\ yields an (regularized) expression for the case $k=1$.

Similarly, for the M\"obius diagram we need the integral:
\eqn\rhss{\eqalign{
\int\limits_0^\infty  dy\ y^\eps\ [E_{2k}(iy-\h)-1]&\cr
&\hskip-30mm=\fc{(2\pi i)^{2k}}{(2k-1)!\ \zeta(2k)}\ (2\pi)^{-1-\eps}\ \Gamma(1+\eps)\ 
\zeta(2+\eps-2k)\ \zeta(1+\eps)\ \Pi(\eps,k)\cr
&\hskip-30mm=\cases{-\h\ 2^{2k-4}\ 
\fc{(2\pi i)^{2k}}{(2k-1)!\ \zeta(2k)}\ (-1)^{k}\ (2\pi)^{1-2k}\ 
\Gamma(2k-1)\ \zeta(2k-1)+\Oc(\eps)\ ,&$k>1\ ,$\cr
\fc{3}{2\pi \eps}+\fc{3\ln 2}{\pi}+\Oc(\eps)\ ,&$k=1$\ .}}}
The projector\foot{In \eqq \following\  it manifests  in the sum over $m$ and $d$ as 
$\h\ [1+(-1)^m]\ [1+(-1)^d]-(-1)^{d+m}$.}   
\eqn\projector{
\Pi(\eps,k)=2^{-1+2k-\eps}+2^{-\eps}-7\cdot 4^{-2+k-\eps}-1}
steams from the $\h$ in the argument of the Eisenstein function.
Thus the effect of the latter is the additional factor $2^{2k-4}$ and a slight modification
in the regularization of the case $k=1$ compared to \rhs.

\appendix{\appB}{Spin--structure sums in the gauged open string partition function}

To perform the even spin--structure sum of \gaugedannc, supplemented
with the relevant piece at $\Oc(B^2)$ order  \expand, the following identity, which
follows from applying the Riemann identity (\cf \eg \STii), is useful:
\eqn\performsum{\eqalign{
\fc{\p^2}{\p z^2}\sum_{\vec\de} s_{\vec\de}&  \lf.
\TH{\de_1}{\de_2}(z,\tau)\ \TH{\de_1}{\de_2}(v^1,\tau)\ \TH{\de_1}{\de_2}(v^2,\tau)\ 
\TH{\de_1}{\de_2}(v^3,\tau)\ri|_{z=0}\cr
&=\h\th_1''[\fc{v^1+ v^2+ v^3}{2}]\ \th_1[\fc{v^1+ v^2- v^3}{2}]\ 
\th_1[\fc{v^1- v^2+ v^3}{2}]\ \th_1[\fc{v^1- v^2- v^3}{2}]\cr
&+\h\th_1[\fc{v^1+ v^2+ v^3}{2}]\ \th_1''[\fc{v^1+ v^2- v^3}{2}]\ 
\th_1[\fc{v^1- v^2+ v^3}{2}]\th_1[\fc{v^1- v^2- v^3}{2}]\cr
&+\h\th_1[\fc{v^1+ v^2+ v^3}{2}]\ \th_1[\fc{v^1+ v^2- v^3}{2}]\ 
\th_1''[\fc{v^1- v^2+ v^3}{2}]\ \th_1[\fc{v^1- v^2- v^3}{2}]\cr
&-\h\th_1[\fc{v^1+ v^2+ v^3}{2}]\ \th_1[\fc{v^1+ v^2- v^3}{2}]\ 
\th_1[\fc{v^1- v^2+ v^3}{2}]\ \th_1''[\fc{v^1- v^2- v^3}{2}]\cr
&-\th_1'[\fc{v^1+ v^2+ v^3}{2}]\ \th_1'[\fc{v^1+ v^2- v^3}{2}]\ 
\th_1[\fc{v^1- v^2+ v^3}{2}]\ \th_1[\fc{v^1- v^2- v^3}{2}]\cr
&+\th_1'[\fc{v^1+ v^2+ v^3}{2}]\ \th_1[\fc{v^1+ v^2- v^3}{2}]\ 
\th_1'[\fc{v^1- v^2+ v^3}{2}]\ \th_1[\fc{v^1- v^2- v^3}{2}]\cr
&+\th_1'[\fc{v^1+ v^2+ v^3}{2}]\ \th_1[\fc{v^1+ v^2- v^3}{2}]\ 
\th_1[\fc{v^1- v^2+ v^3}{2}]\ \th_1'[\fc{v^1- v^2- v^3}{2}]\cr
&+\th_1[\fc{v^1+ v^2+ v^3}{2}]\ \th_1'[\fc{v^1+ v^2- v^3}{2}]\ 
\th_1'[\fc{v^1- v^2+ v^3}{2}]\ \th_1[\fc{v^1- v^2- v^3}{2}]\cr
&-\th_1[\fc{v^1+ v^2+ v^3}{2}]\ \th_1'[\fc{v^1+ v^2- v^3}{2}]\ 
\th_1[\fc{v^1- v^2+ v^3}{2}]\ \th_1'[\fc{v^1- v^2- v^3}{2}]\cr
&-\th_1[\fc{v^1+ v^2+ v^3}{2}]\ \th_1[\fc{v^1+ v^2- v^3}{2}]\ 
\th_1'[\fc{v^1- v^2+ v^3}{2}]\ \th_1'[\fc{v^1- v^2- v^3}{2}]\ .}}
Let us notify the identity $\th_1'(0,\tau)=-2\pi\ \eta(\tau)^3$, which will be used in the 
following.

\subsec{N=1 supersymmetric sector: $v^j\neq 0$ and $\pm v^1\pm v^2\pm v^3=0$}

In this case N=1 supersymmetry is preserved and one--loop corrections to gauge couplings
are generically non--vanishing.
Using \expand\ and \performsum\ the part of \gaugedannc, which is relevant for the 
gauge threshold corrections becomes:
\eqn\Gi{\eqalign{
\lf.\fc{\p^2}{\p B^2}\ \tilde A^{d=3}_{ab}(B)\ri|_{B=0}&=\h q_{a}^2\ I_{ab}\ l\ 
\fc{1}{\eta(\tau)^3}\cr 
&\times\sum_{\vec\de} s_{\vec\de}\ \th''\lf[\de_1\atop\de_2\ri](0,\tau)\ 
\fc{\TH{\de_1}{\de_2}(iv^1,\tau)}{\TH{\h}{\h}(iv^1,\tau)}
\fc{\TH{\de_1}{\de_2}(iv^2,\tau)}{\TH{\h}{\h}(iv^3,\tau)}
\fc{\TH{\de_1}{\de_2}(iv^3,\tau)}{\TH{\h}{\h}(iv^3,\tau)}\cr
&=-\pi\ q_{a}^2\ I_{ab}\ l\ (-1)^{\#}\lf[
\pm\fc{\th_1'(iv^1,\tau)}{\th_1(iv^1,\tau)}\pm\fc{\th_1'(iv^2,\tau)}{\th_1(iv^2,\tau)}\pm
\fc{\th_1'(iv^3,\tau)}{\th_1(iv^3,\tau)}\ri]\ .}}
with $\#$ accounting for the number of minus signs in the combination 
$\pm v^1\pm v^2\pm v^3$ and the modular parameter $\tau=2il$ for the annulus in the closed string 
channel.

\subsec{N=2 supersymmetric sector: $v^i,v^j\neq 0$, $v^k=0$ and $\pm v^i\pm v^j=0$}

In this case N=2 supersymmetry is preserved and again 
one--loop corrections to gauge couplings are generically non--vanishing.
Using \expand and \performsum\ the piece of the gauged open string partition function \gaugedannc, 
needed for the gauge threshold corrections becomes
\eqn\Gii{\eqalign{
\lf.\fc{\p^2}{\p B^2}\ \tilde A^{d=2}_{ab}(B)\ri|_{B=0}&=-\fc{1}{4}\ q_{a}^2\ V_a^k\ 
I^i_{ab}I^j_{ab}\ l\ \fc{1}{\eta(\tau)^6}\ \tilde Z_k(l,T^k,V^k)\cr
&\times \sum_{\vec\de} s_{\vec\de}\ 
\th''\lf[\de_1\atop\de_2\ri](0,\tau)\ \th\lf[\de_1\atop\de_2\ri](0,\tau)
\fc{\TH{\de_1}{\de_2}(iv^i,\tau)}{\TH{\h}{\h}(iv^i,\tau)}
\fc{\TH{\de_1}{\de_2}(iv^j,\tau)}{\TH{\h}{\h}(iv^j,\tau)}\cr
&=-\pi^2\ q_{a}^2\ V_a^k\ I^i_{ab}I^j_{ab}\ l\ \tilde Z_k(l,T^k,V_a^k)\ .}}
For the spin--structure sums in
$\lf.\fc{\p^2}{\p B^2}\tilde M_a(B)\ri|_{B=0}$ and 
$\lf.\fc{\p^2}{\p B^2}\tilde A_{aa'}(B)\ri|_{B=0}$ we obtain similar expressions.

\goodbreak
\appendix{\appC}{$UV$ limits of the gauged open string  partition functions}

In this appendix we derive the $UV$ limit $t\ra 0$ of the expressions {\it Eqs.} 
\gaugedann, \gaugedanni, and \gaugedmoeb. 
We determine this limit from the closed string expressions
\gaugedannc, \gaugedanncc, and \gaugedmoebc, where this limit corresponds
to the $IR$ limit $l\ra \infty$.
We use the product expansion for the $\theta$--functions
\eqn\deftheta{\eqalign{
\TH{\de_1}{\de_2}(z,\tau)&=\sum_{n\in \IZ}e^{\pi i \tau(n+\de_1)^2}\ 
e^{2\pi i(n+\de_1)(z+\de_2)}\cr
&\hskip-7mm=e^{2\pi i \de_1(z+\de_2)}\ q^{\fc{\de_1^2}{2}}\ \prod_{n=1}^\infty (1-q^n)\ 
(1+q^{n+\de_1-\h}\ e^{2\pi i(z+\de_2)})\ (1+q^{n-\de_1-\h}e^{-2\pi i(z+\de_2)})}} 
to derive the limits
\eqn\product{
\lim\limits_{\tau\ra i\infty} \TH{\de_1}{\de_2}(iz,\tau)\ra
\cases{
1+2q^{1/2}\cos(2\pi iz)&, $\ \de_1=0\ \ ,\ \  \de_2=0 $\ ,  \cr
1-2q^{1/2}\cos(2\pi iz)&, $\ \de_1=0\ \ ,\ \  \de_2=\h$\ ,  \cr
2q^{1/8}\cos(\pi iz)   &, $\ \de_1=\h\ \ ,\ \ \de_2=0 $\ ,  \cr 
-2q^{1/8}\sin(\pi iz) &, $\ \de_1=\h\ \ ,\ \ \de_2=\h$\ .   }}
Furthermore we notify $\eta(\tau)\ra q^{1/24}$, and  the useful relations:
\eqn\useful{\eqalign{
\cos(i\pi v^j_a)&=\cosh(\pi v^j_a)=\cos(\phi_a^j)=
\fc{n^j_aR_1^j+m_a^jR_2^j\cos\al^j}{\Vc_a^j}\ ,\cr
\sin(i\pi v_a^j)&=i\sinh(\pi v^j_a)=
\sin(\phi_a^j)=\fc{m_a^jR_2^j}{\Vc_a^j}\sin\al^j\ ,\cr
\cot(i\pi v_a^j)&=-i\coth(\pi v^j_a)=
\cot(\phi_a^j)=\fc{n_a^j R_1^j+m_a^jR_2^j\cos\al^j}{m_a^j R_2^j\sin\al^j}\ ,\cr
\cos(i\pi v^j_{ab})&=\cos(\phi_b^j-\phi_a^j)\cr
&=\fc{1}{\Vc_a^j\Vc_b^j}\ 
\lf[(R_1^j)^2n_a^jn_b^j+(R_2^j)^2m_a^jm_b^j+R_1^jR_2^j(n_a^jm_b^j+n_b^jm_a^j)\cos\al^j\ri]\cr
&=\h\fc{1}{\Vc_a^j\Vc_b^j}\ \fc{T_2^j}{U_2^j}\lf[(n_a^j+m_a^jU^j)(n_b^j+m_b^j\ov U^j)+
(n_a^j+m_a^j\ov U^j)(n_b^j+m_b^jU^j)\ri]\ ,\cr
\sin(i\pi v^j_{ab})&=
\sin(\phi_b^j-\phi_a^j)=R_1^jR_2^j\ \fc{n_a^jm_b^j-n_b^jm_a^j}{\Vc_a^j\Vc_b^j}\ \sin\al^j=
T_2^j\ \fc{I_{ab}^j}{\Vc_a^j\Vc_b^j}\ ,\cr
e^{i(\phi_b^j-\phi_a^j)}&=\fc{R_1^jR_2^j}{\Vc_a^j\Vc_b^j}\ \lf[
\fc{R_1^j}{R_2^j}n_a^jn_b^j+\fc{R_2^j}{R_1^j}m_a^jm_b^j+(n_a^jm_b^j+n_b^jm_a^j)\cos\al^j+
i I_{ab}^j\sin\al^j\ri]\ ,}}
with
\eqn\volumebrane{
\Vc_a^j=\sqrt{\fc{T_2^j}{U_2^j}|n_a^j+m_a^jU^j|^2}=
\sqrt{(n_a^j)^2(R_1^j)^2+(m_a^j)^2(R_2^j)^2+2n_a^jm_a^jR_1^jR_2^j\cos\al^j}} 
describing the volume of brane $a$, which wraps the torus $T_2^j$ 
with wrapping numbers $(n^j_a,m^j_a)$.
The above expressions are written down for an $A$ torus, \ie the lattice basis vector $\vec e_1$ 
is aligned along the $Y^{2i-1}$ axis. However, these expressions may be also used for a B type 
lattice after properly redefining the brane wrapping numbers $(n_a,m_a)$ into 
$(\tilde n_a,\tilde m_a)$.
The latter are determined by the relation of the $A$--lattice basis vectors $\vec e_1,\vec e_2$
to the $B$--lattice basis vectors $\vec e_1',\vec e_2'$.
The same applies for the case of tilted tori.

In the following we shall consider the cases $d=2,3$, appropriate for intersecting $D6$
branes.
For the $R$--sector $\vec\de=(\h,0)$, we obtain in the limit $l\ra \infty$:
\eqn\Ri{\eqalign{
l^{-1}\ \lf.\tilde A^{d=3}_{ab}(B)\ri|_R& \ra I_{ab}\ 
\be_{ab}\cot(\pi \eps_{ab})\prod\limits_{j=1}^3\cot(i\pi v_{ab}^j)\cr
&=\prod\limits_{j=1}^3 \lf[\fc{R_1^j}{R_2^j}n_a^jn_b^j+\fc{R_2^j}{R_1^j}m_a^jm_b^j+
(n_a^jm_b^j+n_b^jm_a^j)\cos\al^j\ri]\ \fc{1}{\sin\al^j}\ ,\cr
l^{-1}\ \lf.\tilde A^{d=2}_{ab}(B)\ri|_R& \ra I^1_{ab}I^2_{ab}\ V_a^3
\ \be_{ab}\cot(\pi \eps_{ab})\prod\limits_{j=1}^2\cot(i\pi v_{ab}^j)\cr
&=V_a^3\prod\limits_{j=1}^2 \lf[\fc{R_1^j}{R_2^j}n_a^jn_b^j+\fc{R_2^j}{R_1^j}m_a^jm_b^j+
(n_a^jm_b^j+n_b^jm_a^j)\cos\al^j\ri]\ \fc{1}{\sin\al^j}.}}

\eqn\Rii{\eqalign{
l^{-1}\ \lf.\tilde A^{d=3}_{aa'}(B)\ri|_R& \ra I_{aa'}\ 
\be_{aa'}\cot(\pi \eps_{aa'})\prod\limits_{j=1}^3\cot(i\pi v_{aa'}^j)\cr
&=(1-\pi^2 q_a q_{a'}B^2)\cr
&\times 
\prod\limits_{j=1}^3 \lf[\fc{R_1^j}{R_2^j}n_a^jn_\ap^j+\fc{R_2^j}{R_1^j}m_a^jm_\ap^j+
(n_a^jm_\ap^j+n_\ap^jm_a^j)\cos\al^j\ri]\ \fc{1}{\sin\al^j}\ ,\cr
l^{-1}\ \lf.\tilde A^{d=2}_{aa'}(B)\ri|_R&\ra
(1-\pi^2 q_a q_{a'}B^2)\ I_{aa'}^1I_{aa'}^2V_a^3\ \cr
&\times \prod\limits_{j=1}^2
\lf[\fc{R_1^j}{R_2^j}n_a^jn_\ap^j+\fc{R_2^j}{R_1^j}m_a^jm_\ap^j+
(n_a^jm_\ap^j+n_\ap^jm_a^j)\cos\al^j\ri]\ \fc{1}{\sin\al^j} }}

\eqn\Riii{\eqalign{
l^{-1}\ \lf.\tilde M^{k;d=3}_{a,\Om\Rc\th^k a}(B)\ri|_R &\ra 
-4\be_a\cot\lf(\fc{\pi \eps_a}{2}\ri)I^k_{a,\Om\Rc\th^k a}\prod_{j=1}^3 2^{\delta_j}
\cot(i\pi v^{k;j}_a)\cr
&=-8iI^k_{a,\Om\Rc\th^k a}\prod\limits_{j=1}^3 2^{\delta_j}\coth(\pi v_a^{k;j})\ ,\cr
l^{-1}\ \lf.\tilde M^{k;d=2}_{a,\Om\Rc\th^k a}(B)\ri|_R &\ra -2^{4-\mu}\be_a 
\cot\lf(\fc{\pi \eps_a}{2}\ri)\ V_{O6_k}^i
\prod\limits_{j=1}^2I^{k;j}_{a,\Om\Rc\th^k a}\cot(i\pi v_{a}^{k;j})\cr
&=2^{5-\mu}\ V_{O6_k}^i\ \prod\limits_{j=1}^2 2^{\delta_j}
I_{a,\Om\Rc\th^k a}^{k;j}\coth(\pi v_a^{k;j})\ .}}
For the NS--sectors $\vec\de=(0,0),\ (0,\h)$ we obtain:
\eqn\NSi{\eqalign{
l^{-1}\ \lf.\tilde A^{d=3}_{ab}(B)\ri|_{NS}& \ra -\fc{i}{2}\ I_{ab}\ 
\fc{1}{\sinh(\pi v_{ab}^1)\sinh(\pi v_{ab}^2)\sinh(\pi v_{ab}^3)}\cr
&\hskip-25mm\times\lf\{\fc{1}{\sqrt{1+\pi^2q_a^2B^2}}+\h\sqrt{1+\pi^2q_a^2B^2}
\ [-1+\cosh(2\pi v_{ab}^1)+\cosh(2\pi v_{ab}^2)+\cosh(2\pi v_{ab}^3)]\ri\},\cr
l^{-1}\ \lf.\tilde A^{d=2}_{ab}(B)\ri|_{NS}& \ra \h I^1_{ab}I^2_{ab}\ V_a^3\ 
\fc{1}{\sinh(\pi v_{ab}^1)\sinh(\pi v_{ab}^2)}\cr
&\times\lf\{\fc{1}{\sqrt{1+\pi^2q_a^2B^2}}+\fc{1}{2}\sqrt{1+\pi^2q_a^2B^2}\ 
[\cosh(2\pi v_{ab}^1)+\cosh(2\pi v_{ab}^2)]\ri\}\ .}}
Furthermore, for gauged open string partition function with both string ends carrying
charges $q_a,q_{a'}$ of the gauge group $G_a$, we determine:
\eqn\NSii{\eqalign{
l^{-1}\ \lf.\tilde A^{d=3}_{aa'}(B)\ri|_{NS}& \ra -\fc{i}{2}\ \ 
\fc{I_{aa'}}{\sinh(\pi v_{aa'}^1)\sinh(\pi v_{aa'}^2)\sinh(\pi v_{aa'}^3)}\cr
&\times\lf\{
\fc{(1-\pi^2q_a q_{a'}B^2)^2}{\sqrt{1+\pi^2q_a^2B^2}\sqrt{1+\pi^2q_{a'}^2B^2}}+
\fc{1}{2}\sqrt{1+\pi^2q_a^2B^2} \sqrt{1+\pi^2q_{a'}^2B^2}\ri.\cr
&\times\lf.[-1+\cosh(2\pi v_{aa'}^1)+\cosh(2\pi v_{aa'}^2)+\cosh(2\pi v_{aa'}^3)]\ri\},\cr
l^{-1}\ \lf.\tilde A^{d=2}_{aa'}(B)\ri|_{NS}& \ra \h\ V_a^3\ 
\fc{I^1_{aa'}I^2_{aa'}}{\sinh(\pi v_{aa'}^1)\sinh(\pi v_{aa'}^2)}\lf\{
\fc{(1-\pi^2q_a q_{a'}B^2)^2}{\sqrt{1+\pi^2q_a^2B^2}\sqrt{1+\pi^2q_{a'}^2B^2}}
\ri.\cr
&+\lf.\fc{1}{2}\sqrt{1+\pi^2q_a^2B^2}\ \sqrt{1+\pi^2q_{a'}^2B^2}
[\cosh(2\pi v_{aa'}^1)+\cosh(2\pi v_{aa'}^2)]\ri\},\cr
l^{-1}\ \lf.\tilde A^{d=0}_{aa'}(B)\ri|_{NS}& \ra \prod\limits_{j=1}^d V_a^j\cr
&\times\lf\{\h\fc{\pi^2\ (q_a+q_{a'})^2\ B^2}
{\sqrt{1+\pi^2q_a^2B^2}\sqrt{1+\pi^2q_{a'}^2B^2}}-
\sqrt{1+\pi^2q_a^2B^2}\sqrt{1+\pi^2q_{a'}^2B^2}\ri\}\ .}}
And for the gauged M\"obius partition function we obtain: 
\eqn\NSiii{\eqalign{
l^{-1}\ \lf.\tilde M^{k;d=3}_{a,\Om\Rc\th^k a}(B)\ri|_{NS}& \ra -4i\ I^k_{a,\Om\Rc\th^k a}
\fc{1}{\sinh(\pi v_{a}^{k;1})\sinh(\pi v_{a}^{k;2})\sinh(\pi v_{a}^{k;3})}\cr
&\hskip-35mm\times\lf\{\fc{1}{\sqrt{1+\pi^2q_a^2B^2}}+\fc{1}{2}\sqrt{1+\pi^2q_a^2B^2}
\ [-1+\cosh(2\pi v_{a}^{k;1})+\cosh(2\pi v_{a}^{k;2})+\cosh(2\pi v_{a}^{k;3})]\ri\},\cr
l^{-1}\ \lf.\tilde M^{k;d=2}_{a,\Om\Rc\th^k a}(B)\ri|_{NS}& \ra -16\ V_{O6_k}^i
I^{k;1}_{a,\Om\Rc\th^k a}\ I^{k;2}_{a,\Om\Rc\th^k a}
\fc{1}{\sinh(\pi v_{a}^{k;1})\sinh(\pi v_{a}^{k;2})}\cr
&\hskip-3mm\times\lf\{\fc{1}{\sqrt{1+\pi^2q_a^2B^2}}+\fc{1}{2}\sqrt{1+\pi^2q_a^2B^2}
\ [\cosh(2\pi v_{a}^{k;1})+\cosh(2\pi v_{a}^{k;2})]\ri\}.}}
Obviously, the $R$--sector ({\it Eqs.} \Ri,\ \Rii\ and \Riii) shows a very simple dependence
on the magnetic field $B$. In fact, only $\lf.\tilde A_{aa'}^d(B)\ri|_{R}$ shows a quadratic order 
in $B$.
However, with taking into account the condition \anomalyfree\ we conclude, that there are no
tadpole contributions in the $R$--sector for anomaly free gauge groups.
This fact is not true for the $NS$--sector, which has generically a functional dependence
on the magnetic field resembling the relevant Schwinger expressions from field theory. 
Note that for the N=2 sector, in the case of $v^1_{ab}=-v^1_{ab}=\pm 1/2$,
this functional behaviour boils down to the expressions for the $\IZ_2$ orientifold  presented in 
\BF. To discuss the possible tadpole contributions in subsection 3.4. we only need from {\it Eqs.} 
\NSi,\ \NSii\ and \NSiii\ the second order parts in $B$, which take the form:
\eqn\NSI{\eqalign{
l^{-1}\ \lf.\tilde A^{d=3}_{ab}(B)\ri|_{NS}&\ra
-i\ I_{ab}\ \lf\{\prod\limits_{j=1}^3\coth(\pi v_{ab}^j)\ri.\cr
&\lf.+\h\pi^2q_a^2B^2\ \lf[\prod\limits_{j=1}^3\coth(\pi v_{ab}^j)-
\prod\limits_{j=1}^3\fc{1}{\sinh(\pi v_{ab}^j)}\ri]\ri\}+\Oc(B^4)\ ,\cr
l^{-1}\ \lf.\tilde A^{d=2}_{ab}(B)\ri|_{NS}&\ra
I^1_{ab}I^2_{ab}\ V_a^3\ \lf\{\prod\limits_{j=1}^2\coth(\pi v_{ab}^j)\ri.\cr
&\lf.+\h\pi^2q_a^2B^2\ \lf[\prod\limits_{j=1}^2\coth(\pi v_{ab}^j)-
\prod\limits_{j=1}^2\fc{1}{\sinh(\pi v_{ab}^j)}\ri]\ri\}+\Oc(B^4)\ .}}
These identities assume the supersymmetry conditions  $v^1_{ab}+v^2_{ab}+v^3_{ab}=0$
for $A_{ab}^{d=3}$ and $v^1_{ab}+v^2_{ab}=0$ for $A_{ab}^{d=2}$, respectively.
Furthermore for the annulus between one brane $a$ from stack $a$ and an other 
brane $a'$ from its mirror:
\eqn\NSII{\eqalign{
l^{-1}\ \lf.\tilde A^{d=3}_{aa'}(B)\ri|_{NS}&\ra
-i I_{aa'}\ \lf\{\prod\limits_{j=1}^3 \coth(\pi v^j_{aa'})\ri.\cr
&\hskip-20mm
\lf.+\h\pi^2B^2\lf[(q_a^2+q_{a'}^2)\ \prod\limits_{j=1}^3 \coth(\pi v^j_{aa'})-
(q_a+q_{a'})^2\ \prod\limits_{j=1}^3 \fc{1}{\sinh(\pi v^j_{aa'})}\ri]\ri\}
+\Oc(B^4)\ ,\cr
l^{-1}\ \lf.\tilde A^{d=2}_{aa'}(B)\ri|_{NS}&\ra
I_{aa'}^1I_{aa'}^2\ V_a^3\ \lf\{\prod\limits_{j=1}^2 \coth(\pi v^j_{aa'})\ri.\cr
&\hskip-20mm
\lf.+\h\pi^2\ B^2\ \lf[(q_a^2+q_{a'}^2)\ 
\prod\limits_{j=1}^2 \coth(\pi v^j_{aa'})-
(q_a+q_{a'})^2\ \prod\limits_{j=1}^2 \fc{1}{\sinh(\pi v^j_{aa'})}\ri]\ri\}
+\Oc(B^4)\ ,\cr
l^{-1}\ \lf.\tilde A^{d=0}_{aa'}(B)\ri|_{NS}&\ra
(-1+B^2\pi^2q_a q_{a'})\ \prod_{j=1}^3V_a^j\cr
&\hskip-20mm=(-1+B^2\pi^2q_a q_{a'})\ \prod\limits_{j=1}^3
\lf[\fc{R_1^j}{R_2^j}(n_a^j)^2+\fc{R_2^j}{R_1^j}(m_a^j)^2
+2n_a^jm_a^j\cos\al^j\ri]\ \fc{1}{\sin\al^j}+\Oc(B^4)\ .}}
Again these identities rely on the supersymmetry conditions  
$v^1_{ab}+v^2_{ab}+v^3_{ab}=0$ for $A_{ab}^{d=3}$ and $v^1_{ab}+v^2_{ab}=0$ 
for $A_{ab}^{d=2}$, respectively.
Finally, for the M\"obius function the lowest expansion in $F$ of \NSiii\ yields
in case of N=1 and N=2 supersymmetry, respectively:
\eqn\NSIII{\eqalign{
l^{-1}\ \lf.\tilde M^{k;d=3}_{a,\Om\Rc\th^k a}(B)\ri|_{NS}& \ra 8i\ I^k_{a,\Om\Rc\th^k a}\ 
\lf\{\prod\limits_{j=1}^3\coth(\pi v_a^{k;j})\ri.\cr
&\lf.+\h\pi^2q_a^2B^2\ \lf[\prod\limits_{j=1}^3\coth(\pi v_a^{k;j})-
\prod\limits_{j=1}^3\fc{1}{\sinh(\pi v_a^{k;j})}\ri]\ri\}+\Oc(B^4)\ ,\cr
l^{-1}\ \lf.\tilde M^{k;d=2}_{a,\Om\Rc\th^k a}(B)\ri|_{NS}& \ra -32\ I_{a,\Om\Rc\th^k a}^{k;1}
I_{a,\Om\Rc\th^k a}^{k;2}\ V^i_{O6_k}\  \lf\{\prod\limits_{j=1}^2\coth(\pi v_a^{k;j}) \ri.\cr
&+\lf.\h\pi^2q_a^2B^2\ \lf[\prod\limits_{j=1}^2\coth(\pi v_a^{k;j})-
\prod\limits_{j=1}^2\fc{1}{\sinh(\pi v_a^{k;j})}\ri]\ri\}+\Oc(B^4)\ .}}

Of course, in the case of supersymmetry, the $R$ and $NS$ vacuum tadpoles
are the same up to a minus sign as a result of vanishing vacuum partition function.
However, this is no longer the case for our open string partition functions depending on
the space--time gauge field $B$. 
For the case $d=3$, describing the N=1 sectors,  the total NS tadpole contributions at order
$B^2$ can be read off from the results \NSI,\ \NSII\ and \NSIII.
Note, that these expressions are the same as the one given 
in \eqqs \closerlook, \closerlooka, and \closerlookm. This is due to the general identity
\eqn\closeri{\eqalign{
&-\coth(\pi v^1)\coth(\pi v^2)\coth(\pi v^3)
+\fc{1}{\sinh(\pi v^1)\sinh(\pi v^2)\sinh(\pi v^3)}\cr
&=\coth(\pi v^1)+\coth(\pi v^2)+\coth(\pi v^3)\ ,}}
valid for the supersymmetric case $v^1+v^2+v^3=0$.
Similar for the N=2 sector contributions $d=2$, derived in
\NSI,\ \NSII\ and \NSIII, we may use the identity
\eqn\closerii{
-\coth(\pi v^1)\coth(\pi v^2)+
\fc{1}{\sinh(\pi v^1)\sinh(\pi v^2)}=1}
for the case $v^1+v^2=0$ to make contact with the UV--divergent expressions from subsection 3.3.

\listrefs
\end